\documentclass[12pt,a4]{article}
\usepackage{soul}
\usepackage{amsthm,amsmath,latexsym,amssymb,amsfonts,amscd,ulem}
\usepackage{graphics,lscape,fancyhdr,array,stmaryrd,euscript}
\usepackage[all]{xy}
\usepackage{empheq}
\usepackage{dsfont}
\usepackage{verbatim}
\usepackage{color,tikz,tikz-cd}
\usetikzlibrary[snakes]
\usepackage{relsize,slashed}
\numberwithin{equation}{section}
\usepackage[colorlinks=true,linkcolor=black,citecolor=black,urlcolor=blue,filecolor=black]{hyperref}
\usepackage{setspace}
 
\usepackage[numbers,sort&compress]{natbib}
\usepackage[nottoc]{tocbibind}

\newcommand{\pl}{\partial}
\newcommand{\besubeqs}{\begin{subequations}}
\newcommand{\esubeqs}{\end{subequations}}

\newcommand{\ga}{\alpha}
\newcommand{\gb}{\beta}

\newcommand{\gad}{{\dot{\alpha}}}

\newcommand{\aI}{{\ensuremath{\mathcal{I}}}}
\newcommand{\aJ}{{\ensuremath{\mathcal{J}}}}
\newcommand{\aK}{{\ensuremath{\mathcal{K}}}}

\newcommand{\fud}[2]{{}^{#1}{}_{#2}\,}

\newcommand{\pd}{\partial}

\newcommand{\cC}{{\cal C}}
\newcommand{\cD}{{\cal D}}

\newcommand{\be}{\begin{equation}}
\newcommand{\ee}{\end{equation}}
\newcommand{\bea}{\begin{eqnarray}}
\newcommand{\eea}{\end{eqnarray}}

\newcommand{\nn}{\nonumber}

\newcommand{\pr}{\partial}
\newcommand{\prd}{\partial \cdot}

\usepackage{soul}

\newcommand{\nbinline}[2][]{
  \ifthenelse { \equal {#1} {} }
    { \def\temp {#2} }  
    { \def\temp {#1} }   
\refstepcounter{todocounter}\todo[color=nbcolor,inline,caption={\textbf{\thetodocounter. NB} \temp}]{\textbf{\thetodocounter. NB:} #2}{}}

\begin{document}

\thispagestyle{empty}
\begin{center}

\vspace*{30pt}
{\LARGE \bf Strange higher-spin topological systems in 3D}

\vspace{30pt}
{Nicolas Boulanger, Andrea Campoleoni, Victor Lekeu and Evgeny Skvortsov}

\vspace{20pt}
{\sl \small
{Service de Physique de l'Univers, Champs et Gravitation,  
Universit\'e de Mons -- UMONS,\\
20 place du Parc, 7000 Mons, Belgium}
}
\vspace{30pt}

\end{center}

\begin{abstract}
Motivated by the generation of action principles from off-shell dualisation, 
we present a general class of free, topological theories in three dimensional
Minkowski spacetime that exhibit higher-spin gauge invariance. 
In the spin-two case, we recover a dual reformulation 
of the triplet system already known, 
while the higher-spin systems that we obtain seem to be new.
They are associated with wild quivers.
We study in which situations these exotic (or strange) higher-spin models 
can be extended to dS${}_3$ and AdS${}_3$ backgrounds, 
revealing that the flat limit of such models, when they exist, 
admits a one-parameter freedom.
Interactions are studied in the simplest higher-spin case featuring 
spin-2 and spin-3 fields.
We then give several higher-spin generalizations of these strange systems. 

\end{abstract}

\vfill
\noindent\texttt{nicolas.boulanger@umons.ac.be, andrea.campoleoni@umons.ac.be,\\
victor.lekeu@umons.ac.be, evgeny.skvortsov@umons.ac.be}

\vspace{25pt}

\newpage

\setcounter{tocdepth}{2}
\tableofcontents

\newpage

\section{Introduction}
\label{sec:Intro}

Recently in \cite{Boulanger:2020yib}, 
two of the authors obtained seemingly new action principles 
for totally-symmetric tensors on a 
three-dimensional Minkowski background  
that enjoy gauge invariance of the higher-spin type.
The procedure for constructing such action
principles followed from the off-shell higher dualisation 
scheme of
\cite{Boulanger:2012df}; see, e.g., 
\cite{Chatzistavrakidis:2020kpx, Boulanger:2022arw} for 
recent applications of this procedure in different contexts.
Due to the importance of higher-spin topological systems in 
three dimensions, see e.g.\ \cite{Henneaux:2010xg, Campoleoni:2010zq, Gaberdiel:2010pz} 
and their subsequent developments, we believe it is important 
to clarify the status of the systems found in \cite{Boulanger:2020yib}, 
which all share the property of describing parity-invariant systems 
in spite of an explicit dependence on the Levi-Civita symbol in their actions. 

One of the motivations of the present paper is therefore to understand 
to which extend these actions are giving new, free topological 
theories in three dimensions that exhibit higher-spin gauge 
invariance. The field content and gauge transformation laws 
severely restrict the possible actions, but field redefinitions
and dualisations may change the form of the actions.

Before turning our attention to topological systems, 
we first consider the simplest 
case of higher dualisation studied in \cite{Boulanger:2020yib}, 
where the original action (hence its dual) describe propagating 
degrees of freedom in three dimensions.  
The starting model is Maxwell theory, and in the dual action 
$S[h_{ab},A_a]$ the fields consist of a vector 
$A_a$ and a traceful rank-two tensor $h_{ab}\,$. 
As we show in section~\ref{sec:propagating},
the dual action $S[h_{ab},A_a]$ of \cite{Boulanger:2020yib} 
reproduces the spin-two triplet system 
already known before \cite{Bengtsson:1986ys, Ouvry:1986dv, Henneaux:1987cp},
after appropriate local field redefinitions.
On the other hand, to the best of our knowledge, the higher-spin topological systems 
presented in \cite{Boulanger:2020yib} never appeared before in the literature.

In the class of 3D topological systems that we consider, 
in section~\ref{sec:topological} we first focus on the simplest 
higher-spin system of \cite{Boulanger:2020yib},
that features symmetric tensor gauge fields of rank two and three, 
and show that it actually belongs to a one-parameter family of action principles, 
with the additional freedom of choosing the relative sign between the canonical 
kinetic terms of the two fields involved. 
By an abuse of terminology, 
we henceforth refer to these systems as spin-3/spin-2 metric-like systems, 
because of their containing rank-two and three symmetric tensor 
gauge fields $h_{ab}$ and $\varphi_{abc}\,$.

Then,
we reformulate the one-parameter family of spin-3/spin-2 metric-like systems 
in first order, frame-like form. 
For this, following the strategy first applied in three dimensions in 
\cite{Blencowe:1988gj}, 
one has to modify the field content in such 
a way that both the spin-2 and \mbox{spin-3} sectors are described off-shell
by a pair of one-forms valued in the spin-1 and spin-2 representations
of the Lorentz group $SO(1,2)$, respectively. 
This is in accordance with the argument explained in \cite{Grigoriev:2020lzu} 
that any (non)interacting topological system in 3D can be formulated as a 
Chern-Simons theory. The resulting actions, however, 
differ from the flat-space spin-2/spin-3 Chern-Simons actions 
of \cite{Afshar:2013vka, Gonzalez:2013oaa}, mainly because in our case the spin-2 
connection is not associated with a $sl(2,\mathbb{R})$ subalgebra of 
a full gauge algebra.
After having performed the frame-like reformulation of the one-parameter 
family of spin-3/spin-2 metric-like systems in flat space,  we 
study their possible non-Abelian Chern-Simons extensions 
and their deformation to the (anti) de Sitter, (A)dS$_3$ background.

We find that the spin-3/spin-2 models in flat space do not admit any 
non-Abelian Chern-Simons deformation.
On the other hand, we discover that, in (A)dS$_3$ background, 
there are very few spin-3/spin-2 models with the same set of fields 
and gauge parameters as in flat space. 
For the flat limit of the action in (A)dS$_3$ to 
exactly reproduce the one-parameter family of actions in flat space,
one has to perform a redefinition of the fields and gauge parameters
before sending the cosmological constant to zero. 
The operations of performing field redefinitions and flat limit do not 
commute. The field redefinition matrices depend on the parameters 
$z\,$ and $\gamma$ that label the family of action principles in flat space 
and the relative sign of the canonical kinetic terms, and on the sign of the 
cosmological constant through a square root. 
As a result, depending on the values of the constants 
$z\,$ and $\gamma\,$ some flat models admit a deformation to  
AdS$_3$, while other flat models admit a deformation to dS$_3$ space only.  

Then, we show how to generalise these spin-2/spin-3 models 
to spins $s>3$ and higher multiplicities of higher-spin fields in the spectrum, 
both in flat and (A)dS$_3$ spaces. These models are related to quivers of the 
wild type, for which a full classification is not available in the general case.
While in flat space this prevents us from classifying all inequivalent models,
in (A)dS$_3$ the semi-simple nature of the spacetime isometry algebra allows 
for a classification, given a spectrum of fields.

\section{Propagating case with a spin two field in 3D}
\label{sec:propagating}

Here we discuss the simplest action $S[h_{ab},A_a]$ studied in 
\cite{Boulanger:2020yib}, that results from the higher dualisation 
of a Maxwell field in three dimensions and that 
features the tensor fields $A_a$ and $h_{ab}\,$, 
of rank one and two, respectively. 
The tensor $h_{ab}$ is symmetric and traceful. 
Differently from all the systems that we will analyse later in this paper, 
the action $S[h_{ab},A_a]$ is not topological: its equations of motion describe 
the propagation of a scalar degree of freedom, as it is the case 
for a Maxwell field in 3D. 
Still, the structure of the action $S[h_{ab},A_a]$ and of the gauge 
transformations that leave it invariant bear similarities with the 
metric-like actions and the gauge transformations of the topological 
higher-spin systems that we will analyse in
section \ref{sec:topological}.
Moreover, the action resulting from the higher dualisation 
of Maxwell's theory belongs to a one-parameter family of inequivalent actions, 
as is the case for its higher-spin counterpart. 

\subsection{From higher dualisation to a family of models}

We first review the higher dualisation of a massless vector 
field in a spacetime of arbitrary dimension presented 
in \cite{Boulanger:2015mka, Boulanger:2020yib}.
The starting point is Maxwell's action, up to boundary terms that we neglect:
\begin{align}
S[A_a] = -\frac{1}{2}\,\int d^n x \left(\partial_a A_b \,\partial^a A^b 
                - \partial_a A^a\,\partial_b A^b \right) .
                \label{Maxwell}
\end{align}
One can then introduce a parent action depending on two fields
$(Y^{ab|c}\,,P_{ab})\,$ that have no symmetries under permutations 
of their indices, apart from the antisymmetry $Y^{ab|c}=-Y^{ba|c}\,$: 
\begin{align}
    S[Y^{ab|c},P_{ab}] = \int d^n x \left( P_{ab}\,\partial_cY^{ca|b}
    -\tfrac{1}{2}\,P_{ab}\,P^{ab} + \tfrac{1}{2}\,P_a{}^a\,P_b{}^b \right) .
\end{align}
Extremising with respect to $Y^{ab|c}$ imposes $P_{ab}=\partial_a A_b$
that, when substituted inside the parent action, reproduces Maxwell's action
\eqref{Maxwell}. On the other hand, $P_{ab}$ is an auxiliary field. Solving 
for it inside the parent action yields the dual action
\begin{align}
    S[Y^{ab|c}] = \int d^n x \left( \tfrac{1}{2}\,\partial_c Y^{ca|b}\,\partial_dY^d{}_{a|b}
    - \tfrac{1}{2(n-1)}\,\partial_a\,Y^{ab|}{}_b\,\,\partial_c\,Y^{cd|}{}_d \right) ,
\end{align}
invariant under the gauge transformations
\begin{align}
    \delta Y^{ab|}{}_c = \delta^{[a}{}_c\,\partial^{b]}\epsilon 
    + \partial_d \psi^{abd}{}_c\;,\qquad \psi^{abd}{}_c = \psi^{[abd]}{}_c\;.
\end{align}
Here and below, indices enclosed by a pair of (square) brackets denote a 
(anti)symmetrisation with strength one, where dividing by the number of terms 
in the (anti)symmetrisation is understood.
The $GL(n,\mathbb{R})$ irreducible decomposition of 
$Y^{ab|}{}_c$ reads $Y^{ab|}{}_c = X^{ab}{}_c + 2\,\delta_c{}^{[a} Z^{b]}\,$, 
where $X^{ab}{}_a\equiv 0\,$.
In dimension $n=3\,$, the above decomposition amounts to 
\begin{align}
 Y^{ab|}{}_c = \varepsilon^{abc}\,h_{cd} + 2\,\delta_c{}^{[a} Z^{b]}\,,
 \qquad h_{ab} = h_{ba}\;.
\label{YintermsofhandZ}
\end{align}
Sticking to the dimension $n=3\,$, 
the dual action in terms of $h_{ab}$ and $Z_a$ reads
\begin{align}
    S[h_{ab},Z_a] = \int d^3x \left( -\tfrac{1}{2}\,\partial_ah_{bc}\,\partial^ah^{bc}
    +\tfrac{1}{2}\,\partial_ah_{bc}\,\partial^{b}h^{ac}
    +\tfrac{1}{2}\,\varepsilon^{bcd}\,F_{cd}\,\partial^ah_{ab}
    +\tfrac{1}{4}\,F^{ab}\,F_{ab} \right) ,
    \label{dualactionhZ}
\end{align}
where $F_{ab}=2\,\partial_{[a}Z_{b]}\,$. The above action is invariant under 
the gauge transformations
\begin{align}
    \delta h_{ab} = 2\,\partial_{(a}\xi_{b)}\;,
    \qquad \delta Z_{a} = \partial_a \epsilon 
    + \varepsilon_{abc}\,\partial^b\xi^c\;.
    \label{gaugesymmetryhZ}
\end{align}
It is also possible to dualise the vector field $Z_a$ into a scalar field $\phi\,$, 
following the standard procedure that we will discuss in section~\ref{sec:dual-spin1}.
After dualisation, the action reads \cite{Boulanger:2020yib}
\begin{align}
     S[h_{ab},\varphi]=\int d^3x
    \left[-\tfrac{1}{2}\,\partial_ah_{bc}\,\partial^ah^{bc}
    + \partial_ah^{ab}\,\partial^ch_{bc}
    +2\,\partial_a\varphi\,(\partial^a\varphi+\partial_b h^{ab})\right]\;,
    \label{eq:2.8}
\end{align}
where the field $h_{ab}$ transforms as in \eqref{gaugesymmetryhZ} and 
the scalar $\varphi$ transforms as $\delta \varphi = -\partial^a\xi_a\,$.
After the field redefinition $\varphi = \frac{1}{2}\,(\phi + h)$
that combines the trace of the field $h_{ab}$
with the new scalar field $\phi$, one obtains the equivalent action
\begin{align}
\label{dualactionhphi}
    S[h_{ab},\phi]=\int d^3x
    &[-\tfrac{1}{2}\,\partial_ah_{bc}\,\partial^ah^{bc}
    +\tfrac{1}{2}\,\partial_ah\,\partial^ah - \partial_ah\,\partial_bh^{ab}
    + \partial_ah^{ab}\,\partial^ch_{bc}
    \nonumber \\
    &\qquad + \tfrac{1}{2}\,\partial_a\phi\,\partial^a\phi
    + \partial_a\phi(\partial_bh^{ab}-\partial^ah)]\;,
\end{align}
where the scalar field $\phi$ does not transform and the field $h_{ab}$ 
still transforms as in \eqref{gaugesymmetryhZ}. 
The terms quadratic in the first derivative of 
$h_{ab}$ reproduce the (massless) Fierz-Pauli Lagrangian.
Extremising the action with respect to both fields and combining the 
field equation for $\phi$ with the trace of the field equation for 
$h_{ab}$ yields the following set of equations:
\begin{align}
    \Box\phi = 0\;,\quad 
    \Box h - \partial^{a}\partial^bh_{ab} = 0\;,
    \quad  R_{ab} := \partial_{ab}h - 2\,\partial_{(a}\partial^c h_{b)c}
    +\Box h_{ab} = \partial_a\partial_b \phi\;.
    \label{eomdualactionhphi}
\end{align}
The first two equations taken alone would lead to a doubling of 
degrees of freedom corresponding to two scalars propagating in 3 dimensions. 
However, the ``wrong" relative sign for the kinetic terms of the 
two fields inside the action \eqref{dualactionhZ} or its dual
\eqref{dualactionhphi} is responsible for the third equation in 
\eqref{eomdualactionhphi}, which identifies the curvatures of 
the two fields and reduces the degrees of freedom to a single 
scalar, in agreement with the starting point for a Maxwell field 
in 3D. 
Indeed, by construction, the higher dualisation procedure does not 
change the number of degrees of freedom. See \cite{Boulanger:2020yib}
for more discussions and generalisations to higher dimensions. 

The action \eqref{dualactionhZ} resulting from the higher 
dualisation of a Maxwell field in 3D is, in fact, a member 
of the following one-parameter family of actions,
\begin{align} 
S_{1}[h_{ab},Z_a] = \frac{1}{2} \int d^3x \, \big( 
& - \partial_a h_{bc} \partial^a h^{bc} + 
(\alpha+2)\, \partial\cdot h_a \partial\cdot h^a - (\alpha+1)\, 
\partial^a h \left[\,  2\, \partial\cdot h_a - \partial_a h \,\right] 
\nonumber \\
& - \tfrac{\ga}{2}\, F_{ab} F^{ab} - \ga\, \varepsilon_{abc}\, \pr\cdot h^a F^{bc} 
\big) \, , \label{flat-action}
\end{align}
which is invariant under the gauge transformations \eqref{gaugesymmetryhZ}.
The action \eqref{dualactionhZ} is recovered for $\alpha = -1\,$.
Notice that, although all these actions exhibit the 
antisymmetric Levi-Civita symbol, 
they are invariant under both parity and time-reversal transformations, under which 
the fields transform as $A_a \mapsto A_a$ and 
$h_{ab} \mapsto - h_{ab}\,$. 
The sign flip in the latter transformation can be understood recalling 
that the field $h_{ab}$ first appeared in \eqref{YintermsofhandZ} contracted 
with an antisymmetric 3D symbol. The same transformations can then be 
postulated for all members of the one-parameter family of action principles. 

The family of actions \eqref{flat-action} provides a simple example of 
couplings between 
free fields in Minkowski space induced  by $\varepsilon$-terms, 
that we further explore in section \ref{sec:topological}.  
In this case, for $\alpha = 0\,$ the vector field disappears 
from the action \eqref{flat-action},   
that reduces to the Fierz-Pauli action in 3D Minkowski space.
This leads to a discontinuity in the number of propagating degrees of freedom, 
since the higher dualisation procedure preserves the number of propagating degrees 
of freedom, while the Fierz-Pauli action does not propagate any degrees of 
freedom in three dimensions. 

\subsection{Dualisation of the spin-1 field and link with 
a spin-2 triplet}\label{sec:dual-spin1}

As anticipated in the previous section, since the action 
\eqref{flat-action} only depends on $A_a$ via its field strength, 
the vector field can be dualised into a scalar \cite{Boulanger:2020yib}. 
To this end, one considers the antisymmetric tensor $F_{ab}$ 
as an independent field in place of the curl of $Z_a\,$ 
that appears in $S_1[h_{ab},Z_a]\,$ (see Eq. \eqref{flat-action}),
and one constructs the following action:
\be
S_{\textrm{parent}}[h_{ab},F_{ab},\varphi] 
= S_1[h_{ab},F_{ab}] + \alpha \int d^3x\, 
\varepsilon_{abc}\, \varphi \, \pr^a F^{bc} \, ,
\ee
where we fixed the normalisation 
so as to simplify some of the ensuing formulae.
Extremising the new action with respect to $\varphi$ one recovers the Bianchi identity for the field strength as an equation of motion:
\be
\frac{\delta S_{\textrm{parent}}}{\delta \varphi} = 0 
\quad \Rightarrow \quad 
\pr_{[a} F_{bc]} = 0
\quad \Rightarrow \quad
F_{ab} = \pr_{a} Z_b - \pr_b Z_a \, .
\ee
Extremising the parent action with respect to $F_{ab}$ one obtains instead
\be
\frac{\delta S_{\textrm{parent}}}{\delta F^{ab}} = 0 
\quad \Rightarrow \quad 
F_{ab} = -\, \varepsilon_{abc} \left( 2\, \pr^c \varphi + \prd h^c \right) ,
\ee
and substituting this algebraic relation into the action gives
\be \label{flat-action-dual}
\begin{split}
S_0[h_{ab}, \varphi] = \frac{1}{2} \int d^3x \, 
\big( & - \pr_a h_{bc} \pr^a h^{bc} + 2\, \pr\cdot h_a \pr\cdot h^a 
- (\ga+1)\, \pr^a h \left[\,  2\, \pr\cdot h_a - \pr_a h \,\right] \\
& - 4\ga \left[\, \pr_a \varphi\, \pr^a \varphi - \varphi\, \prd\prd h \,\right] 
\big) \, .
\end{split}
\ee
This action is invariant under
\be
\delta h_{ab} = 2\, \pr_{(a} \xi_{b)} \, , 
\qquad
\delta \varphi = -\, \prd \xi \, .
\ee
In fact, it is easy to check that the above action 
\eqref{flat-action-dual} is gauge invariant in arbitrary dimension.
For $\alpha = -1$ one recovers the action (2.28) of \cite{Boulanger:2020yib}
which we reproduced above in \eqref{eq:2.8}.
For the same value of $\alpha$, the action also corresponds to that 
of a spin-2 triplet \cite{Bengtsson:1986ys, Ouvry:1986dv, Henneaux:1987cp} 
after the elimination of the field with an algebraic equations of motion. 
The action of a spin-2 triplet, reviewed, e.g., 
in~\cite{Francia:2002pt, Sagnotti:2003qa}, indeed reads
\be \label{triplet}
S_{\textrm{triplet}} = \int d^n x \left( - \tfrac{1}{2}\, \pr_a h_{bc} \pr^a h^{bc} + 2\, \prd h_a \, \cC^a + 2\, \prd \cC\,\cD + \pr_a \cD\, \pr^a \cD - \cC_a \cC^a \right)
\ee
and it is invariant under
\be
\delta h_{ab} = 2\, \pr_{(a} \xi_{b)} \, , 
\qquad
\delta \cC_\mu = \Box \xi_a \, , 
\qquad
\delta \cD = \prd \xi \, .
\ee
The equation of motion for $\cC_a$ is algebraic:
\be
\cC_a = \prd h_a - \pr_a \cD \, .
\ee
Substituting it into the action \eqref{triplet} one gets
\be \label{tripletD}
S_{\textrm{triplet}} = \int d^n x \left( - \tfrac{1}{2}\, \pr_a h_{bc} 
\pr^a h^{bc} +  \pr\cdot h_a \pr\cdot h^a + 2 \left[\,  
\pr_a \cD\, \pr^a \cD + \cD\, \prd\prd h \,\right]\right) ,
\ee
that is, when $D = 3$, the action \eqref{flat-action-dual} with 
$\cD = - \varphi$ and $\alpha = -1$.

Alternatively, as pointed out in \cite{Campoleoni:2012th}, 
the action \eqref{tripletD} 
can also be obtained starting from the Maxwell-like action
\be \label{maxwell-like}
S_{\textrm{M-L}} = \int d^n x \left( - \tfrac{1}{2}\, \pr_a h_{bc} \pr^a h^{bc} +  \pr\cdot h_a \pr\cdot h^a \right) ,
\ee
with a traceful $h_{ab}$.\footnote{Considering the same action with a traceless 
$h_{ab}$ along the lines of \cite{Skvortsov:2007kz} gives instead an action 
equivalent to the Fierz-Pauli one. 
A similar pattern applies to higher-spin fields: Maxwell-like actions for 
traceless fields are equivalent to Fronsdal ones \cite{Skvortsov:2007kz}, 
while the same actions for traceful fields are equivalent to higher-spin 
triplet systems \cite{Campoleoni:2012th}. Reducible spectra with less 
propagating fields can also be obtained by imposing the vanishing of only 
some traces of the fields \cite{Francia:2016weg}.} 
This action is invariant under
\be
\delta h_{ab} = 2\, \pr_{(a} \xi_{b)} 
\quad \textrm{with} \quad
\prd\xi = 0 \, .
\ee
The differential constraint on the gauge parameter can however 
be eliminated via the Stueckelberg shift
\be
h_{ab} \to h_{ab} - 2\, \pr_{(a} \theta_{b)} \, ,
\ee
where the new field transforms as $\delta \theta_a = \xi_a$.
The resulting action can only depend on $\theta_a$ via its divergence: 
introducing the field \mbox{$\cD = \prd \theta$} gives back the 
action \eqref{tripletD}.

In conclusion, the  indecomposable system obtained from the higher dualisation 
of the Maxwell action in three dimensions corresponds to the dualisation of the already known 
indecomposable system given by the triplet \cite{Bengtsson:1986ys, Ouvry:1986dv, Henneaux:1987cp}, in its simplified version involving only two fields 
\cite{Francia:2002pt, Sagnotti:2003qa, Campoleoni:2012th}. 
The dualisation substituting the triplet's scalar with a vector 
is, obviously, only possible in three dimensions.
On the other hand, the one-parameter family of actions \eqref{flat-action-dual} 
and, consequently, the action \eqref{tripletD} for $\alpha=-1$,
can be formulated in any space-time dimensions.

\subsection{Deformation to (A)dS}

The triplet system can be deformed to (A)dS \cite{Bengtsson:1990un, Buchbinder:2001bs, Sagnotti:2003qa, Bonelli:2003zu, Barnich:2006pc, Campoleoni:2012th}; 
similarly, the action \eqref{flat-action-dual} involving a scalar besides 
the rank-two field admits a deformation to (A)dS for any value of the 
parameter $\alpha$. The action
\be
\begin{split}
S_0 = \frac{1}{2} \int d^3x\, \big( & - \nabla_{\!a} h_{bc} \nabla^a h^{bc} 
+ 2\, \nabla\cdot h_a \nabla\cdot h^a - (\ga+1)\, \nabla^a h 
\left[\,  2\, \nabla\cdot h_a - \nabla_{\!a} h \,\right] 
\\
& - 4\ga \left[\, \nabla_{\!a} \varphi\, \nabla^a \varphi - \varphi\, 
\nabla\cdot\nabla\cdot h \,\right] - 2\,\sigma\lambda^2 \left[ h_{ab} h^{ab} 
+ \alpha\, h^2 - 4\alpha\, \varphi^2 \right] \big) \, ,
\end{split}
\ee
where $\nabla$ denotes the (A)dS covariant derivative while we parameterize the 
cosmological constant as $\Lambda = - \sigma \lambda^2$, is indeed invariant under
\be
\delta h_{ab} = 2\, \nabla_{\!(a} \xi_{b)} \, , 
\qquad
\delta \varphi = -\, \nabla\cdot \xi \, .
\ee

On the other hand, it is not possible to preserve a deformation of the gauge 
symmetry \eqref{gaugesymmetryhZ} of the action \eqref{flat-action} involving 
a vector besides the rank-two field. The most general action giving back 
\eqref{flat-action} in the flat, $\lambda \to 0$ limit is
\begin{align}
S = \frac{1}{2} \int\! d^3x\, \big( & - \nabla_{\!a} h_{bc} \nabla^a h^{bc} 
+ (\alpha+2)\, \nabla\cdot h_a \nabla\cdot h^a - (\alpha+1) \left[\,  
2\, \nabla\cdot h_a \nabla^a h - \nabla_{\!a} h \nabla^a h \,\right] 
\nn \\
& - \tfrac{\alpha}{2}\, F_{ab} F^{ab} - \alpha\, \varepsilon_{abc} \nabla\cdot 
h^a F^{bc} + \lambda\, a_1\, Z^a \nabla\cdot h_a + \lambda\, a_2\, h 
\nabla\cdot Z \label{general-ansatz-AdS} \\[5pt]
& + \sigma \lambda^2 \left[\, m_1^2\, h_{ab} h^{ab} +m_2^2\, h^2 + m_3^2\, 
Z_a Z^a \,\right] \big) \, , \nn
\end{align}
and one can consider gauge transformations of the type
\be
\delta h_{ab} = 2\, \nabla_{\!(a} \xi_{b)} + \sigma \lambda\, k_1\, g_{ab} 
\epsilon  \, , \qquad
\delta A_a = \pr_a \epsilon + \varepsilon_{abc} \pr^b \xi^c + \sigma \lambda\, k_2\, 
\xi_a \, .
\ee
Still, it is not possible to preserve the gauge symmetries generated by 
$\xi_a$ and $\epsilon$ for any choice of the coefficients.\footnote{Choosing 
$a_1 = a_2 = 0$, $m_1^2 = 2(\alpha-1)$, $m_2^2 = -2\alpha$, $m_3^2 = 4\alpha$ 
together with $k_1 = k_2 = 0$ allows one to preserve the gauge symmetry generated 
by $\xi_a$, while the one generated by $\epsilon$ is broken by the 
non-vanishing mass-like term for the vector.}
This result anticipates some subtlelties in the deformation to (A)dS of the 
action principles directly given or suggested by the higher dualisation procedure 
that we shall encounter in the following sections, although we shall discuss this 
issue mainly in the frame-like reformulation of our new models. 

\section{Family of spin-2/spin-3 topological systems}
\label{sec:topological}

In this section, we analyse in details the simplest exotic 
model with higher-spin gauge symmetry, a model that one obtains from
the higher dualisation of a massless spin-2 field in three-dimensional
Minkowski spacetime. We first show that it actually belongs
to a one-parameter family of inequivalent exotic models in flat space,
whose spectrum of fields consists 
of the pair $(h_{ab},\varphi_{abc})$ of traceful, symmetric tensors. 
We also deform these flat-space spin-2/spin-3 models to 
the (A)dS$_3$ background and show that the one-parameter freedom disappears.
In other words, there exists only a discrete number of spin-2/spin-3 models 
in (A)dS$_3\,$.
Retrospectively, this means that there is a one-parameter freedom 
in taking the flat limit of the models in (A)dS$_3\,$, 
at least at the level of the equations of motion.

\subsection{A family of models}
\label{sec:family}

It turns out that the action found in \cite{Boulanger:2020yib} from the dualization of 
the Fierz-Pauli action in three dimensions is a member of a the following 
family of actions for the traceful, symmetric tensors $h_{ab}$ and 
$\varphi_{abc}\,$,\footnote{The symbols $h$ and $\varphi_a\,$ denote, 
respectively, the 
trace of the tensors $h_{bc}$ and $\varphi_{abc}\,$.} 
\begin{align}
    S[\varphi_{abc},h_{ab}] =\frac{1}{2} \int d^3\!x\, \Big(&
    \;a_0 \, \partial_a\varphi_{bcd}\,\partial^a\varphi^{bcd}+
    a_1 \, \partial^a\varphi^b\,\partial^c\varphi_{abc}
    + a_2 \, \partial_a\varphi^{abc}\,\partial^d\varphi_{bcd}
    \nonumber \\
    & +a_3 \,\partial_a \varphi_b\,\partial^a \varphi^b + a_4 \,\partial_a 
    \varphi^a\,\partial^b \varphi_b
    \nonumber \\
    & + b_0 \,\partial_a h_{bc}\, \partial^a h^{bc}
    + b_1 \,\partial_a h\, \partial^a h
    + b_2 \,\partial^a h_{ab}\, \partial_c h^{bc}
    + b_3 \,\partial^a h\, \partial_c h_a{}^c
    \nonumber \\
    & + c_1 \,\varepsilon_{pqr}\,\partial^a h_{a}{}^p\,
    \partial^q\varphi^r 
    + c_2 \,\varepsilon_{apq}\,\partial^b h^{ac}\,
    \partial^p\varphi^q{}_{bc}\; \Big) \, . \label{eq:action23family}
\end{align}
These actions are invariant under gauge transformations of the 
form 
\besubeqs\label{coupledgaugetrans}
\begin{align}
    \delta \varphi_{abc} &= 3\,\partial_{(a}\xi_{bc)} 
    - 3x\,\varepsilon_{(a}{}^{pq}\,\eta_{bc)} \pd_p \epsilon_q \;, \\[5pt]
    \delta h_{ab} &= 2\, \pd_{(a}\epsilon_{b)} 
    -2z\,\varepsilon_{pq(a} \partial^p \xi^q{}_{b)} \;,
\end{align}
\esubeqs
where the gauge parameter $\xi_{ab}$ is symmetric and traceless and 
the parameters $x$ and $z$ are fixed by the requirement of gauge invariance, 
leading to an interesting one-parameter family of models. This is done in 
several steps:
\begin{itemize}
    \item First of all, we assume that the spectrum of fields indeed 
    involves a genuine rank-3
symmetric tensor $\varphi_{abc}\,$, with its traceless part 
appearing in the action: this means that the parameters $a_0$, $a_1$, $a_2$ 
and $c_2$ cannot all vanish. Under this assumption, we find that 
$b_0$ is never zero: we can therefore choose to fix it to $\pm 1/2$ 
by rescaling the $h_{ab}$ field and possibly by flipping the sign of the whole action. 
We will write $\gamma$ for this sign choice: thus, $b_0 = \gamma/2\,$.
\item Next, one finds that $x=0$ if and only if $z=0$: in that case,
there is no mixing in the gauge transformations. 
As a result, one gets $c_1 = c_2 = 0$, i.e.~no terms in the Lagrangian mixing 
the two fields. 
The action then reduces to the sum (or difference) 
of the Fronsdal action for spin $3$ 
and the Fierz-Pauli action for spin $2$ with some arbitrary relative 
sign $\gamma\,$. 
In what follows, we will therefore assume $x\neq 0$ and $z \neq 0\,$, 
implying that at least $c_1$ or $c_2$ is nonvanishing: there is a genuine 
mixing in the action between the two fields.
\item We then use the freedom of rescaling the $\varphi_{abc}$ field. 
The generic case is $a_0\neq 0$: we can then fix $a_0 = -1$ (possibly by again 
flipping the sign of the whole action), and $\gamma$ is the relative 
sign between the $\varphi_{abc}$ and $h_{ab}$ kinetic terms. 
All the other parameters are then fixed in terms of $\gamma$ and 
the parameter $z$: from the requirement of gauge invariance, one finds
\begin{equation}\label{eq:xofz}
    x = -\frac{2 \gamma  z}{9 \left(3 \gamma  z^2-2\right)}\;,
\end{equation}
and
\besubeqs\label{eq:abcvalues}
\begin{align}
    a_0 &= -1\, , \quad a_1 = 7 \gamma  z^2-6\, , \quad a_2 = 3-2 \gamma  z^2\, , \\
    a_3 &= -\frac{1}{d}\left( 49 z^4-75 \gamma  z^2+27\right)\, , \quad a_4 = -\frac{1}{4d} \left( 172 z^4-195 \gamma  z^2+54 \right)\, ,\\
    b_0 &= \frac{\gamma}{2}\, , \quad b_1 = -\frac{\gamma}{2d} \left( 8 \gamma  z^2-9\right)\, , \quad b_2 = -\frac{3\gamma}{d} \left(4 \gamma  z^2-3\right) , \\
    b_3 &= \frac{\gamma}{d} \left( 8 \gamma  z^2-9 \right)\, , \quad c_1 = -\frac{2 \gamma  z}{d} \left(14 \gamma  z^2-9\right)\, , \quad c_2 = 2 \gamma  z\, ,
\end{align}
\esubeqs
where the denominator $d$ appearing in several terms is $d = 16 \gamma  z^2-9\,$. 
This is a genuine one-parameter family of inequivalent actions, as all the freedom of 
field rescalings has been used up. The action of \cite{Boulanger:2020yib} describing 
a higher dualisation of the Fierz-Pauli action is recovered for 
$\gamma = + 1$ and $z = -1\,$ (hence $x=2/9$). 
As observed in \cite{Boulanger:2020yib}, 
a wrong relative sign between the kinetic terms is a characteristic of 
actions obtained by the higher-dualisation procedure. 
Note that for $\gamma = +1$, there are values for the parameter $z$ where 
\eqref{eq:xofz} or \eqref{eq:abcvalues} diverge. The spectrum changes at those values: 
the case $z = \pm 3/4$ (where $d=0$) corresponds, after multiplying the action by a 
global factor of $d$, to the case $a_0 = a_1 = a_2 = c_2 = 0$ where the traceless part 
of $\varphi_{abc}$ drops out of the action. Similarly, the case $z = \pm \sqrt{2/3}$ 
where \eqref{eq:xofz} diverges corresponds to removing the usual gauge transformation 
of the spin two field, i.e., the first term in $\delta h_{ab}\,$. 
We therefore exclude these cases.
\item We finally discuss the remaining exotic case $a_0 = 0$, where the usual 
kinetic term $\partial_a\varphi_{bcd}\,\partial^a\varphi^{bcd}$ for the spin 3 
field is absent. This is an isolated point: we fix the normalisation 
of $\varphi_{abc}$ by $a_2 = -1$, and the solution reads
\besubeqs\label{eq:a00}
\begin{align}
    a_0 &= 0\, , \quad a_1 = \frac{7}{2}\, , \quad a_2 = -1 \, , \quad a_3 = -\frac{49}{32} \, , \quad a_4 = -\frac{43}{32} \, , \\
    b_0 &= \frac{1}{2} \, , \quad b_1 = -\frac{1}{4} \, , \quad b_2 = -\frac{3}{4} \, , \quad b_3 = \frac{1}{2} \, , \\
    c_1 &= -\frac{7}{4\sqrt{2}} \, , \quad c_2 = \sqrt{2} \, , \\
    x &= -\frac{2\sqrt{2}}{27} \, , \quad z = \frac{1}{\sqrt{2}}\, .
\end{align}
\esubeqs
\end{itemize}

\subsection{Dualisation of the spin-2 field}

Inside the above one-parameter action, 
in the generic case with $a_0=-1\,$, the spin-2 field $h_{ab}$ 
appears only through its curl
\begin{align}
    \omega_{abc}(h) := \partial_a h_{bc} - \partial_b h_{ac} \;.
\end{align}
Indeed, one finds that
\begin{align}
    S[\varphi_{abc},h_{ab}] = \frac{1}{2}\int d^3x \Big[&
    -\partial_a\varphi_{bcd}\,\partial^a\varphi^{bcd}+
    a_1\,\partial^a\varphi^b\,\partial^c\varphi_{abc}
    +a_2\,\partial_a\varphi^{abc}\,\partial^d\varphi_{bcd}
    \nonumber \\
    & +a_3\,\partial_a \varphi_b\,\partial^a \varphi^b
    +a_4\,\partial_a \varphi^a\,\partial_b \varphi^b
    + \tfrac{\gamma}{4}\,\omega^{abc}(h)\,\omega_{abc}(h)
    + \tfrac{\beta}{2d}\,\omega^{ab\,}{}_b(h)\,\omega_{ac\,}{}^c(h)
    \nonumber \\
    & +\,\varepsilon_{abc}\,\omega^{ab\,}{}_d(h)
    \left(\tfrac{\mu}{d}\,\partial^d\varphi^c + \nu\,\partial^e\varphi_e{}^{cd}\right) 
    \Big] \;, \label{eq:action23bis}
\end{align}
where the constants $a_1\,$, $a_2\,$, $a_3$, and $a_4$ take the same 
values as in \eqref{eq:abcvalues} while
\begin{align}
    \beta = 9\gamma - 8 z^2\;,\quad d = 16\gamma z^2 - 9\;,\quad 
    \mu = z(14z^2-9\gamma)\;, \quad \nu = -\gamma\,z\;.
\end{align}

As a result, one can dualise the spin-2 field $h_{ab}$ and trade it for 
a vector field $A_a$, in analogy with what we did for the spin-1 field in 
section~\ref{sec:dual-spin1}. 
This is achieved by introducing the parent action in the usual way, 
with the coupling $\varepsilon_{abc}\,\omega^{ab}{}_d\,\partial^cA^d\,$, 
where now $\omega_{abc}$ is an independent field satisfying the
following algebraic symmetries:
\begin{align}
    \omega_{bac} \equiv - \omega_{abc}\;,\qquad \omega_{[abc]}\equiv 0\;. 
\end{align}
The field $\omega_{abc}$ is auxiliary. One can solve for it in terms of the 
fields $\varphi_{abc}$ and $A_a$ by using its own field equations. 
Upon substituting the corresponding expression of the auxiliary field $\omega_{abc}$
inside the parent action $S[\varphi_{abc},\omega_{abc},A_a]\,$, one finds the action
\begin{align}\label{action31}
S[\varphi_{abc},A_a] = \int d^3x\,\Big[& 
\,\tfrac{8z^2}{9}\,\partial^a A^b\,\partial_a A_b
+k_1\,\partial^a A^b\,\partial_b A_a
\nonumber \\
& +k_2\,\partial_a\varphi_b\,\partial^b A^a 
+k_3\,\partial_a\varphi_b\,\partial^a A^b
-2z\,\partial^c\varphi_{abc}\,\partial^a A^b
\nonumber \\
& +k_4\,\partial_a\varphi^a\,\partial_b\varphi^b
+k_5\,\partial_b\varphi_c\,\partial^b\varphi^c
+\tfrac{3}{2}\,\partial_a\varphi^{abc}\,\partial^d\varphi_{bcd}
\nonumber \\
&+k_6\,\partial^c\varphi^b\,\partial^d\varphi_{bcd}
-\tfrac{1}{2}\,\partial_d\varphi_{abc}\,\partial^d\varphi^{abc}
\Big]\;
\end{align}
for some definite values of the six parameters $\{k_i\}_{i=1,\ldots,6}\,$ 
that are functions of the parameters $z$ and $\gamma\,$, 
and that we will not need to specify here. 

The above action $S[\varphi_{abc},A_a]$ is invariant under 
\begin{subequations}\label{newgauge}
\begin{align}
    \delta \varphi_{abc} &= 3\,\partial_{(a}\xi_{bc)} 
    - 3\,x\,\varepsilon_{(a}{}^{pq}\,\eta_{bc)} \pd_p \epsilon_q\;,
    \quad x = -\tfrac{2 \gamma  z}{9 \left(3 \gamma  z^2-2\right)}\;,
    \label{gauge1} \\[5pt]
    \delta A_{a} &= \alpha\,\partial^b\xi_{ab} 
    + \beta\,\epsilon_{abc} \,\partial^b \epsilon^c \;,
    \quad
    \alpha = z\,\tfrac{56\,\gamma\, z^2-27}{32\,z^2-18\,\gamma}\;,\qquad 
    \beta = \tfrac{280\,\gamma\,z^4-423\,z^2+162\,\gamma}{864\,z^4-1062\,\gamma\,z^2+324}\;. \label{gauge2}
\end{align}
\end{subequations}
We note that the gauge transformation of $A_a$ is not proportional to
the gauge transformation of the trace $\varphi_{a}\,$ and that 
there is no real value for $z$ such that the parameter 
$\beta$ would vanish.
The vector field is thus independent of the spin-3 field.
One can perform the  field redefinition
\begin{align}
    \phi_{abc} &:= \varphi_{abc} + \zeta\;\eta_{(ab}A_{c)}\;,\qquad
    \zeta = -\tfrac {12\left( 48\,\gamma\, z^5 - 59\, z^3 + 
     18\,\gamma \,z \right)} {840\, z^6 - 1829\,\gamma \,z^4 + 
   1332 \,z^2-324\,\gamma }
\end{align}
that leads to a transformation law where the vector gauge parameter $\epsilon_a\,$
drops out: 
\begin{align}
    \delta \phi_{abc} &= 3\,\partial_{(a}\xi_{bc)} 
    - \tau\;\eta_{(ab}\partial^d\xi_{c)d}\;,
\end{align}
for some value of the parameter $\tau\,$ we do not need to display here.
The point is that the vector field $A_a$ still transforms with the 
parameter $\epsilon_a\,$, therefore it is not possible to have a set 
of independent fields $\{A_a,\varphi_{abc}\}\,$ 
both of which being inert under the $\epsilon_a$ gauge transformations,
showing that the action principle \eqref{action31} cannot be recast 
into a spin-$3$ triplet system. 
Moreover, a spin-$3$ triplet propagates a scalar
degree of freedom even in $3D$, while any member of our family of 
actions is topological. 
We shall make this manifest in section~\ref{sec:firstorderreformulation} 
by showing that the action \eqref{eq:action23bis} can be rewritten in 
Chern-Simons-like form.

Therefore, to the best of our knowledge, with the action \eqref{eq:action23bis}
we have a genuinely new action principle for a topological system involving a spin-3 
and a spin-2 gauge fields or, equivalently, a spin-3 and a spin-1 fields, if one 
chooses to 
dualise the spin-2 field into a spin-1 field, as we have done in this section. In the 
latter case, one can also notice the absence of the Levi-Civita symbol in the action 
\eqref{action31}. Moreover, the Levi-Civita symbol and the gauge parameter 
$\epsilon_a$ only enter the gauge transformations \eqref{newgauge} via the 
combination $\epsilon_{abc} \,\partial^b \epsilon^c$ that can be traded for a 
divergenceless vector. In analogy with what we observed for
the action \eqref{flat-action-dual}, this suggests the option to define an 
action with the same field content and similar gauge transformations also 
in a Minkowski background of arbitrary dimension. Indeed, the action
\begin{equation}
\begin{split}
S[\varphi_{abc}, A_a] = \frac{1}{2} \int d^n x \Big( & -\partial_a\varphi_{bcd}\,\partial^a\varphi^{bcd}+
    \tilde{a}_1\,\partial^a\varphi^b\,\partial^c\varphi_{abc}
    +\tilde{a}_2\,\partial_a\varphi^{abc}\,\partial^d\varphi_{bcd}
    \\
    & + \tilde{a}_3\,\partial_a \varphi_b\,\partial^a \varphi^b
    +\tilde{a}_4\,\partial_a \varphi^a\,\partial_b \varphi^b \\
    & + n(n+1) \partial_a A_b \partial^a A^b + \tilde{b}_1\, \partial_a A_b \partial^b A^a \\
    & + \tilde{c}_1\, \partial^c \varphi_{abc} \partial^a A^b + \tilde{c}_2\, \partial_a \varphi_b \partial^a A^b + \tilde{c}_3\, \partial_a \varphi_b \partial^b A^a \Big)
\end{split}
\end{equation}
is invariant under
\begin{align}
    \delta \varphi_{abc} = 3\left( \partial_{(a}\widehat{\xi}_{bc)} 
    + \eta_{(ab} \Lambda_{c)} \right) , \quad 
    \delta A_{a} = \sqrt{3} \left( \alpha\,\prd\widehat{\xi}_{a} 
    + \tfrac{(n+2)\alpha \pm 2}{2}\, \Lambda_a \right) , \quad
    \prd \Lambda = 0 \;,
\end{align}
for any value of the space-time dimension $n$ provided that
\besubeqs\label{eq:tildea00}
\begin{align}
    \tilde{a}_1 & = - 3 \left( 2 \pm n\,\alpha \right) , \quad 
    \tilde{a}_2 = 3 \, , \quad 
    \tilde{a}_3 = \tfrac{3}{4} \left(n\,\alpha \left((n+1)\alpha \pm 4 \right) +4 \right) , \\[5pt]
    \tilde{a}_4 & = \tfrac{3}{8} \left(n\,\alpha \left((n-2)\alpha \pm 4 \right) +4 \right) , \quad
    \tilde{b}_1 = \tfrac{1}{2}\,n(n-2) \, , \\[5pt]
    \tilde{c}_1 &= \pm 2\sqrt{3}\,n \, , \quad 
    \tilde{c}_2 = - \sqrt{3}\,n \left( (n+1) \alpha \pm 2 \right) , \quad
    \tilde{c}_3 = -\tfrac{\sqrt{3}\,n}{2} \left( (n-2)\alpha \pm 2 \right) .
\end{align}
\esubeqs
We fixed the normalisation of the fields by conventionally fixing the coefficients 
in front of the terms $\partial_a\varphi_{bcd}\,\partial^a\varphi^{bcd}$ 
and $\partial_a A_b \partial^a A^b$, taking into account that gauge 
invariance requires them to have opposite sign.
Notice that we obtained a one-parameter family of actions and that the field 
$A_b$ cannot be gauged away because the gauge parameter $\Lambda_a$ is 
divergenceless, while $A_a$ is an arbitrary vector. To the best of our knowledge, 
the action \eqref{action31}, which has the same field content of a spin-3 
triplet but displays a different gauge symmetry, was never studied before 
it will be interesting to analyze its spectrum for $n > 3$.

In the following we will reformulate the new action 
principle \eqref{eq:action23bis} in an Abelian Chern-Simons-like form.
After we have done it, we will be able to study its possible non-Abelian 
deformations and to generalise it to many new topological systems 
in both flat and (A)dS$_3$ backgrounds.

\subsection{First-order reformulation and non-Abelian deformation}
\label{sec:firstorderreformulation}

We now investigate the first-order formulation of the 
family of models \eqref{eq:action23family}, in terms of one-forms 
$(e^a, \omega^{ab})$ for the spin-2 field and $(E^{ab}, \Omega^{ab,c})$ 
for the spin-3 field, in agreement with the strategy first 
developed in \cite{Vasiliev:1980as, Blencowe:1988gj}. 
In particular, we will recover the gauge 
transformations \eqref{coupledgaugetrans} for the fields 
$h_{ab}$ and $\varphi_{abc}$ after a Lorentz-like partial gauge fixing, 
and exhibit an Abelian Chern-Simons-like action for these models, 
in accordance with 
the general discussion in \cite{Grigoriev:2020lzu}.
We recall that these models are defined around the Minkowski three-dimensional 
background. In Section \ref{sec:AdSdeformation} 
we will consider deformations to the (A)dS$_3$ background.

\paragraph{Gauge transformations.} A general ansatz for the variations 
of the one-forms $e^{a} = e_c{}^a \, {\rm d}x^c$ and $E^{ab} = E_c{}^{ab} \, 
{\rm d}x^c$ is
\besubeqs\label{eq:gaugevielbeins}
\begin{align}
    \delta E_{a,bc} &= \pd_a \xi_{bc} - \alpha_{bc,a} + x \,(\eta_{bc} 
    \tilde{\Lambda}_a - 3 \eta_{a(b}\tilde{\Lambda}_{c)}) \;,\\
    \delta e_{a,b} &= \pd_a \epsilon_b - \Lambda_{ab} + 2z\, \tilde{\alpha}_{ab}\; ,
\end{align}
\esubeqs
where
\begin{equation}\label{eq:dualparameters}
\tilde{\alpha}_{ab} = \tfrac{1}{2}\,\varepsilon_{apq}\,\alpha_b{}^{p,q}\;,
\qquad 
\tilde{\Lambda}_{a} = \tfrac{1}{2}\,\varepsilon_{abc}\,\Lambda^{bc}\;.
\end{equation}
Here, $E_{a,bc}$ is symmetric and traceless in $b, c$ with no other symmetry
involving the index $a\,$. 
The parameter $\alpha_{bc,a}$ is a traceless hook in the symmetric convention: 
\begin{align}
\alpha_{bc,a} &= \alpha_{(bc),a}\;,\qquad \alpha_{(bc,a)} = 0\;,
\qquad \eta^{bc}\alpha_{bc,a} = 0 = \eta^{ab}\alpha_{bc,a}\;.  
\end{align}
The Lorentz parameter $\Lambda_{ab}$ is antisymmetric. 
These parameters are those appearing in the gauge transformations of the connections 
$\omega^{bc} = \omega_a{}^{bc}\, {\rm d} x^a$ and $\Omega^{bc,d} = \Omega_a{}^{bc,d}\, {\rm d} x^a$:
\begin{align}\label{eq:gaugeconnections}
    \delta \omega_a{}^{bc} = \pd_a \Lambda^{bc}\, , \quad \delta \Omega_a{}^{bc,d} 
    = \pd_a \alpha^{bc,d}\, .
\end{align}

From the transformations \eqref{eq:gaugevielbeins}, 
it is clear that we can use $\alpha$ and $\Lambda$ to gauge-fix to zero
the corresponding components of the frame-like fields $e^a$ and $E^{ab}\,$, 
i.e., the traceless hook part of $E_{a,bc}$ and the antisymmetric part 
of $e_{a,b}\,$. 
One calls such a gauge the Lorentz-like gauge. 
Residual gauge transformations then have to satisfy 
$\delta E_{a,bc}\big|_{\text{traceless hook}} = 0$ and $\delta e_{[a,b]} = 0\,$: 
this gives
\begin{align}
    \alpha_{bc,a}^\text{res.} &= \pd_a \xi_{bc} \big|_{\text{traceless hook}} 
    = \pd_a \xi_{bc} - \pd_{(a}\xi_{bc)} 
    + \frac{1}{3}\left(\eta_{bc}\,\pd\!\cdot\!\xi_a  
    - \eta_{a(b}\,\pd\!\cdot\!\xi_{c)}\right) \, ,\\
    \Lambda_{ab}^\text{res.} &= \pd_{[a} \epsilon_{b]}\, .
\end{align}
Notice that there is no entanglement of gauge parameters here: 
indeed, $\Lambda$ only appears through pure trace 
terms in $\delta E_{a,bc}$ so is not involved in its traceless part. 
Similarly, because 
$\alpha$ is traceless, we have $\tilde{\alpha}_{[ab]}=0$ identically and 
therefore $\alpha$ does not appear in $\delta e_{[a,b]}\,$.

After the Lorentz-like gauge-fixing, the Fronsdal and Fierz-Pauli fields
\begin{equation}
    \varphi_{abc} := 3\, E_{(a,bc)}\, , \qquad h_{ab} := 2\, e_{(a,b)}
\end{equation}
transform as
\begin{align}
    \delta \varphi_{abc} = 3\,\partial_{(a}\xi_{bc)} 
    - 3 x \,\varepsilon_{pq(a}\,\eta_{bc)} \pd^p \epsilon^q\;,
    \quad \delta h_{ab} = 2\, \pd_{(a}\epsilon_{b)} 
    - 2 z \,\varepsilon_{pq(a} \partial^p \xi^q{}_{b)} \;,
\end{align}
making contact with the original gauge transformations 
\eqref{coupledgaugetrans}.

In differential form notation (that we will use from now on), the gauge 
transformations \eqref{eq:gaugevielbeins}--\eqref{eq:gaugeconnections} 
read\footnote{Repeated 
covariant or contravariant indices are implicitly symmetrised with strength one.}
\besubeqs\label{eq:gaugespin3}
\begin{align}
    \delta E^{aa} &= {\rm d}\xi^{aa} - h_b\,\alpha^{aa,b} 
    + x\, ( \eta^{aa}\,h^b\,\tilde{\Lambda}_b - 3 h^a \tilde{\Lambda}^a) \nonumber\\
    &= {\rm d}\xi^{aa} +\tfrac{4}{3}\, h_b\,\varepsilon^{abc}\,\tilde\alpha_c{}^a
    - 3x\, (h^a \tilde{\Lambda}^a - \tfrac{1}{3}\,\eta^{aa}\,h^b\,\tilde{\Lambda}_b)\, ,
    \\[5pt]
    \delta \Omega^{aa} &= {\rm d}\tilde{\alpha}^{aa}
\end{align}
\esubeqs
for the spin 3 sector, and
\besubeqs\label{eq:gaugespin2}
\begin{align}
    \delta e^a &=  {\rm d}\xi^{a} + h_b\,\Lambda^{ab} 
    + 2 z\, h_b\,\tilde{\alpha}^{ab}
    \nonumber \\
    &= {\rm d}\xi^{a} - \varepsilon^{abc}\,h_b\,\tilde{\Lambda}_c 
    +2 z\, h_b\,\tilde\alpha^{ab}
    \;,
    \\[5pt]
    \delta \omega^{a} &= {\rm d}\tilde{\Lambda}^{a}
\end{align}
\esubeqs
for the spin 2 sector. 
We have dualised the connection one-forms, similarly to \eqref{eq:dualparameters}:
$\omega_a = \tfrac{1}{2}\,\varepsilon_{abc}\,\omega^{bc}\;$, 
$\Omega_{ab} = \tfrac{1}{2}\,\varepsilon_{apq}\,\Omega_b{}^{p,q}\;$.
These objects carry less indices and we will use them exclusively in what follows.
The one-forms $h^a$ are the background dreibeins 
for Minkowski space: e.g., in Cartesian coordinates they read 
$h^a = \delta^a{}_b \,{\rm d}x^b$.

\paragraph{First-order action for the strange topological system.}

The first-order action invariant under the gauge transformations 
\eqref{eq:gaugespin3}--\eqref{eq:gaugespin2} is
\begin{align}\label{FirstOrderAction2-3}
    S[e^a,\omega^a,E^{aa},\Omega^{aa}] = 
    \int_{M_3} & \Big[
    \omega_a 
    ({\rm d}e^a -\frac{1}{2} \,\varepsilon^{apq} \,h_p\, \omega_q ) 
    + 2z \;\omega_a h_b\, \Omega^{ab} \nonumber\\
    &\quad + \frac{2z}{3x} \; \Omega_{aa} 
    ( {\rm d}E^{aa} + \frac{2}{3}\,
    \varepsilon^{apq} \,h_p\, \Omega_q{}^a ) 
    \Big]\;.
\end{align}
This action can be rewritten in the form
\begin{equation}\label{eq:actionfirstordercurvatures}
    S[e^a,\omega^a,E^{aa},\Omega^{aa}] = 
    \frac{1}{2} \int_{M_3} \Big[ \omega_a R^a(e) + e_a R^a(\omega) 
    + \frac{2z}{3x} \left( \Omega_{ab} R^{ab}(E) + E_{ab} R^{ab}(\Omega) \right) 
    \Big]\, ,
\end{equation}
where the invariant curvatures read
\begin{subequations}
\begin{align}
    R^{aa}(E) &= {\rm d}E^{aa} +\tfrac{4}{3}\, h_p\,\varepsilon^{pqa}\,\Omega_q{}^a
    - 3x\, (h^a \omega^a - \tfrac{1}{3}\,\eta^{aa}\,h^b\,\omega_b)\;,
    &R^{aa}(\Omega) &= {\rm d}\Omega^{aa}\;,
    \\[5pt]
    R^a(e) &= {\rm d}e^{a} - \varepsilon^{abc}\,h_b\,\omega_c 
    + 2z\, h_b\,\Omega^{ab} \;,
    &R^a(\omega) &= {\rm d}\omega^{a}\;.
\end{align}
\end{subequations}
These curvatures satisfy the Bianchi identities
\begin{subequations}
\begin{align}
    0 &\equiv{\rm d} R^{aa}(E) +\tfrac{4}{3}\, h_p\,\varepsilon^{pqa}\,R_q{}^a(\Omega)
    -3x\, (h^a R^a(\omega) - \tfrac{1}{3}\,\eta^{aa}\,h^b\,R_b(\omega))\;,
    &0 &\equiv {\rm d} R^{aa}(\Omega)\;,\\[5pt]
    0 &\equiv {\rm d} R^a(e) -\; \varepsilon^{abc}\,h_b\,R_c(\omega)\;
    + 2z\, h_b\,R^{ab}(\Omega)
    \;,
    &0 &\equiv {\rm d} R^a(\omega)\;.
\end{align}
\end{subequations}
The relative factor $2z/3x$ between the spin two and spin three parts 
of the action \eqref{eq:actionfirstordercurvatures} is necessary for gauge invariance.
The field equations obtained from the above action simply read
\begin{align}
    R^a(e)=0\;,\quad R^a(\omega) =0 \;,\quad R^{aa}(E)=0 \;,\quad R^{aa}(\Omega)=0 \;.
    \label{eq:eomflat}
\end{align}

As it is clear from the form \eqref{FirstOrderAction2-3} of the action, the 
connections $\omega^a$ and $\Omega^{aa}$ are auxiliary fields: they 
can be expressed in terms of the frame-like fields $e^a$ and $E^{aa}$
by solving their field equations algebraically. 
The first-order action principle, upon expressing the auxiliary fields 
in terms of $e^a$ and $E^{aa}$, then gives a second-order action principle for the 
latter fields which is \eqref{eq:action23family} 
in the case $x, z \neq 0$.

In particular, in the special case where $z=-1$ and $\gamma=+1\,$, therefore
$x=2/9\,$, we have shown that the Abelian Chern-Simons-like action 
\eqref{FirstOrderAction2-3} reproduces the metric-like
action obtained in \cite{Boulanger:2020yib} 
by performing an off-shell higher-dualisation of 
three-dimensional linearised gravity around Minkowski background, 
as expected from \cite{Grigoriev:2020lzu}.

\paragraph{Non-Abelian deformations.}
Now that we have reformulated the spin-3/spin-2 systems in first-order form, 
we are ready to study their non-Abelian deformations. 
It turns out that, unfortunately, there is none. 

We search for a nonlinear extension of 
\eqref{eq:gaugespin3}--\eqref{eq:gaugespin2} and 
\eqref{FirstOrderAction2-3} in the form 
\begin{align}
    S[e^a,\omega^a,E^{aa},\Omega^{aa}] 
    = \int_{M_3} {\rm Tr}(\tfrac{1}{2}\,A{\rm d}A + \tfrac{1}{3}\,A^3)\;. 
\end{align}
From the structure of the action \eqref{FirstOrderAction2-3}, 
it is clear that the relevant connection 1-form is 
\begin{align}
    A = \omega^a \,J_a + (h^a+e^a) \,P_a + \Omega^{ab} \,J_{ab} + E^{ab} \,P_{ab} \;. 
\end{align}
It is also clear that, up to a normalisation, 
we have the standard Killing form \cite{Afshar:2013vka, Gonzalez:2013oaa}:
\begin{align}
{\rm Tr}( J_a  P_b) = \eta_{ab}\;,\qquad 
{\rm Tr}(J_{ab}  P_{cd}) = \frac{z}{3x} \,(\eta_{ac}\eta_{bd} + \eta_{ad}\eta_{bc}  
- \tfrac{2}{3}\,\eta_{ab}\eta_{cd} )\;.
\end{align}
We search for a nonlinear extension of the action \eqref{FirstOrderAction2-3}, 
which entails finding a non-Abelian algebra for the $8+8$ generators 
$\{J_a,J_{ab},P_a,P_{ab}\}\,$. From the linearised action and the gauge 
transformations laws 
\eqref{eq:gaugespin3}--\eqref{eq:gaugespin2}, 
we can already read off some of the commutation relations, 
those that imply the generators of the background 
connection $A_0 = h^a\,P_a\,$:
\begin{align}
    [P_a,J_b] &= - \,\varepsilon_{abc}\,P^c -3x\,P_{ab} \;,
    &[P_a , P_b] = 0\;,
    \\
    [P_a,J_{bc}] &= 2z\left( \eta_{a(b}\,P_{c)} 
    -\tfrac{1}{3}\,\eta_{bc}\,P_a \right) 
    + \tfrac{4}{3}\,\varepsilon_{a(b}{}^m\,P_{c)m} \;,
    &[P_a,P_{bc}] = 0 \;.
\end{align}
This is our initial datum.
We must now parametrise all the other commutators 
and constrain them via the Jacobi identities. 
The idea of the proof is to write down the most general 
Ansatz for the other commutators and check whether the Jacobi 
identities are satisfied. 
There are a priori twenty Jacobi identities to be checked.
We find that at least one of these identities cannot be satisfied, 
for all nonzero values of the parameters $x$ and $z\,$. 
There is therefore no non-Abelian deformation of the theory. 

Notice that, had we found a Lie algebra, the resulting non-Abelian 
Chern-Simons action would have been an exotic higher-spin 
extension of 3D gravity, in the sense that the spin-2 sector 
would not have been a consistent truncation of the full theory, 
differently from the higher-spin theories considered in, e.g., 
\cite{Blencowe:1988gj, Henneaux:2010xg, Campoleoni:2010zq, Afshar:2013vka, Gonzalez:2013oaa}. 
In other words, the generators 
$\{P_a,J_a\}$ would not have formed a subalgebra, as is clear from 
the commutator $[P_a,J_b] = - \,\varepsilon_{abc}\,P^c -3x\,P_{ab}\,$, 
since the parameter $x$ is nonzero. 

\subsection{Systems in (A)dS backgrounds}
\label{sec:AdSdeformation}

In this section, 
we first classify the most general gauge-invariant, first-order field 
equations for the 
fields $(e^a,\omega^a,E^{aa},\Omega^{aa})$ 
in (A)dS$_3\,$ that can be cast as zero-curvature conditions. 
We find that, in (A)dS$_3\,$, it is always possible to perform 
field redefinitions\footnote{Such redefinitions are not available 
in Minkowski space since the fields have different dimensions; 
in (A)dS$_3$, a dimensionful parameter (the cosmological constant) 
is available to resolve the mismatch.} 
within the spin-2 sector $(e^a,\omega^a)$ 
and the spin-3 sector $(E^{aa},\Omega^{aa})\,$ 
in such a way as to produce seven inequivalent models, 
six of which are defined in AdS$_3$ and one in dS$_3$.
An action admitting these 
field equations upon variation can then be constructed. 
This is in sharp contrast with the continuous one parameter (and a sign) 
family of actions \eqref{eq:actionfirstordercurvatures} 
and equations \eqref{eq:eomflat} in flat space. 

We then consider the flat limit of the field 
equations and actions in (A)dS$_3\,$ and investigate whether 
one can reproduce the family \eqref{eq:actionfirstordercurvatures} and \eqref{eq:eomflat} of models in flat space.
For this, one first has to perform field redefinitions in both 
spin sectors $(e^a,\omega^a)$ and $(E^{aa},\Omega^{aa})\,$, 
bringing in the free parameter $z\,$, before sending the cosmological 
constant to zero. 
We find that this is always possible for the field equations, 
while the limit of the action can be defined only when the free parameter $z$ 
assumes a finite set of numerical values.
Equivalently, it means that the family of flat space actions 
\eqref{eq:actionfirstordercurvatures}
admits a deformation to (A)dS$_3$ only when the free parameter 
$z$ assumes some very specific values: only a finite number of 
members of the family admit a deformation to (A)dS$_3\,$.

Since we are dealing with topological systems, 
the difference between the Minkowski and (A)dS$_3$ backgrounds reflects 
the crucial difference in the representation theory of the corresponding 
isometry algebras. The finite-dimensional, non-unitary 
representations of the Poincar\'e algebra are much more numerous than 
those of the isometry algebras of the (A)dS$_3$ backgrounds 
\cite{Paneitz:1984, Paneitz:1985ajb, Lenczewski:1986, Gruber:1986, Jakobsen2011}.

\paragraph{Conventions.} 
Going to (A)dS$_3$ background amounts to considering the Lorentz-covariant 
derivative one-form valued operator $\nabla=h^a \,\nabla_a$ such 
that 
\begin{align} \label{eq:cov-der}
    \nabla^2 V^a = -\sigma \lambda^2\; h^a\,h_b\,V^b\;.
\end{align}
The sign parameter $\sigma$ is such that $\sigma = 1$ corresponds 
to anti-de Sitter and $\sigma = -1$ to de Sitter, 
and the $h^a$ are now the vielbeins of (A)dS.
In these conventions, the one-forms $h^a$ have a dimension of length, 
while $\lambda$ has the dimension of mass. 
The cosmological constant is  
$\Lambda=-\sigma\,\lambda^2\,$. 

\paragraph{Gauge transformations and curvatures.}
The most general gauge transformations 
and corresponding gauge-invariant curvatures for 
the spectrum of fields considered in the previous section are
\besubeqs
\begin{align}
\delta e^a &= \nabla\xi^{a} + \lambda x_1\, \varepsilon^{abc}\,h_b\,\xi_c +x_2\, \varepsilon^{abc}\,h_b\,\tilde{\Lambda}_c +\lambda x_3 \, h_b \xi^{ab} + x_4\, h_b\,\tilde\alpha^{ab} \;,\\[5pt]
\delta \omega^a &= \nabla \tilde\Lambda^a + \lambda^2 x_5\,\varepsilon^{abc}h_b \xi_c + \lambda x_6\, \varepsilon^{abc}\,h_b\,\tilde{\Lambda}_c
+ \lambda^2 x_7\,h_b\,\xi^{ab} + \lambda x_8\,h_b\,\tilde\alpha^{ab}\;,\\[5pt]
\delta E^{aa} &= \nabla\xi^{aa} +\lambda x_9 \, (h^a \xi^{a} - \tfrac{1}{3}\,\eta^{aa}\,h^b\,\xi_b) +x_{10}\, (h^a \tilde{\Lambda}^a - \tfrac{1}{3}\,\eta^{aa}\,h^b\,\tilde{\Lambda}_b)\nonumber\\
&\qquad+\lambda x_{11} \, h_b\,\varepsilon^{abc}\,\xi_c{}^a
+x_{12} \, h_b\,\varepsilon^{abc}\,\tilde\alpha_c{}^a \;,\\[5pt]
\delta \Omega^{aa} &= \nabla\tilde{\alpha}^{aa} +\lambda^2 x_{13} \, (h^a \xi^{a} - \tfrac{1}{3}\,\eta^{aa}\,h^b\,\xi_b) +\lambda x_{14} \, (h^a \tilde{\Lambda}^a - \tfrac{1}{3}\,\eta^{aa}\,h^b\,\tilde{\Lambda}_b)\nonumber\\
&\qquad+\lambda^2 x_{15} \, h_b\,\varepsilon^{abc}\,\xi_c{}^a
+\lambda x_{16} \, h_b\,\varepsilon^{abc}\,\tilde\alpha_c{}^a\;,
\end{align}
\esubeqs
and
\besubeqs
\begin{align}
R^a(e) &= \nabla e^{a} + \lambda x_1\, \varepsilon^{abc}\,h_b\,e_c +x_2 \varepsilon^{abc}\,h_b\,\omega_c +\lambda x_3 \, h_b E^{ab} + x_4\, h_b\,\Omega^{ab}\; , \\[5pt]
R^a(\omega) &= \nabla \omega^a + \lambda^2 x_5\,\varepsilon^{abc} h_b e_c + \lambda x_6\, \varepsilon^{abc}\,h_b\,\omega_c
+ \lambda^2 x_7\,h_b\,E^{ab} + \lambda x_8\,h_b\,\Omega^{ab}\;,\\[5pt]
R^{aa}(E) &= \nabla E^{aa} +\lambda x_9 \, (h^a e^{a} - \tfrac{1}{3}\,\eta^{aa}\,h^b\,e_b) +x_{10}\, (h^a \omega^a - \tfrac{1}{3}\,\eta^{aa}\,h^b\,\omega_b)\nonumber\\
&\qquad+\lambda x_{11} \, h_b\,\varepsilon^{abc}\,E_c{}^a
+x_{12} \, h_b\,\varepsilon^{abc}\,\Omega_c{}^a \;,\\[5pt]
R^{aa}(\Omega) &= \nabla\Omega^{aa} +\lambda^2 x_{13} \, (h^a e^{a} - \tfrac{1}{3}\,\eta^{aa}\,h^b\,e_b) +\lambda x_{14} \, (h^a \omega^a - \tfrac{1}{3}\,\eta^{aa}\,h^b\,\omega_b)\nonumber\\
&\qquad+\lambda^2 x_{15} \, h_b\,\varepsilon^{abc}\,E_c{}^a
+\lambda x_{16} \, h_b\,\varepsilon^{abc}\,\Omega_c{}^a\; .
\end{align} 
\esubeqs
The Bianchi identities take the same form: for example,
\begin{equation}
    0 \equiv \nabla R^a(e) + \lambda x_1\, \varepsilon^{abc}\,h_b\,R_c(e) + x_2 \,\varepsilon^{abc}\,h_b\,R_a(\omega) +\lambda x_3 \, h_b R^{ab}(E) +x_4\, h_b\,R^{ab}(\Omega)\, .
\end{equation}
It is useful to rewrite the gauge transformations 
in matrix form
\besubeqs
\label{eq:gauge23matrix}
\begin{align}
    \delta\begin{pmatrix}
        \lambda e^a \\  \omega^a
    \end{pmatrix} &= \nabla \begin{pmatrix}
        \lambda \xi^a \\  \tilde{\Lambda}^a
    \end{pmatrix} + \lambda A\, \varepsilon^{abc} h_b \begin{pmatrix}
        \lambda \xi_c \\  \tilde{\Lambda}_c
    \end{pmatrix} + \lambda B \, h_b \begin{pmatrix}
        \lambda \xi^{ab} \\ \tilde{\alpha}^{ab}
    \end{pmatrix}\;,\\
    \delta\begin{pmatrix}
        \lambda E^{aa} \\ \Omega^{aa}
    \end{pmatrix} &= \nabla \begin{pmatrix}
        \lambda \xi^{aa} \\ \tilde{\alpha}^{aa}
    \end{pmatrix} + \lambda C\, \left( h^a\delta^a_b - \tfrac{1}{3} \eta^{aa}h_b\right) \begin{pmatrix}
        \lambda \xi^b \\ \tilde{\Lambda}^b
    \end{pmatrix} + \lambda D\, \varepsilon^{abc} h_b \begin{pmatrix}
        \lambda \xi_c{}^{a} \\ \tilde{\alpha}_c{}^{a}
    \end{pmatrix}\;,
\end{align}
\esubeqs
with the matrices, only involving dimensionless coefficients, explicitly given by
\begin{equation}
    A = \begin{pmatrix}
        x_1 & x_2 \\ x_5 & x_6
    \end{pmatrix}\, , \quad
    B = \begin{pmatrix}
        x_3 & x_4 \\ x_7 & x_8
    \end{pmatrix}\, , \quad
    C = \begin{pmatrix}
        x_9 & x_{10} \\ x_{13} & x_{14}
    \end{pmatrix}\, , \quad
    D = \begin{pmatrix}
        x_{11} & x_{12} \\ x_{15} & x_{16}
    \end{pmatrix}\, . 
\end{equation}
The requirement of gauge invariance of the curvatures 
gives sixteen quadratic equations on the parameters $x_i\,$. 
In matrix form, they read
\besubeqs \label{eq:matrixeqs}
\begin{align}
    A^2 - \tfrac{5}{6}\, B C &= \sigma I\, , \\
    \tfrac{1}{2}\, D^2 + C B &= 2 \sigma I\, , \\
    -\tfrac{3}{2}\, D C + C A &= 0\, , \\
    -\tfrac{3}{2}\, B D + A B &= 0\, .
\end{align}
\esubeqs
We look for solutions that mix the spin-2 and spin-3 sectors, 
implying that the matrices $B$ and $C$ cannot simultaneously vanish.

We can of course redefine the fields and gauge parameters in each sector,
\besubeqs\label{eq:fieldredefMN}
\begin{align}
    \begin{pmatrix}
        \lambda e^a \\  \omega^a
    \end{pmatrix} = M \begin{pmatrix}
        \lambda e'^a \\  \omega'^a
    \end{pmatrix}\, , \quad \begin{pmatrix}
        \lambda \xi^a \\ \tilde{\Lambda}^a
    \end{pmatrix} = M \begin{pmatrix}
        \lambda \xi'^a \\  \tilde{\Lambda}'^a
    \end{pmatrix} \;,\\
    \begin{pmatrix}
        \lambda E^{aa} \\ \Omega^{aa}
    \end{pmatrix} = N \begin{pmatrix}
        \lambda E'^{aa} \\ \Omega'^{aa}
    \end{pmatrix} \, , \quad
    \begin{pmatrix}
        \lambda \xi^{aa} \\ \tilde{\alpha}^{aa}
    \end{pmatrix} = N \begin{pmatrix}
        \lambda \xi'^{aa} \\ \tilde{\alpha}'^{aa}
    \end{pmatrix}\;,
\end{align}
\esubeqs
with $M, N$ arbitrary $GL(2,\mathbb{R})$ matrices. 
Then, for the primed fields and gauge parameters, 
the gauge transformations take the same form 
\eqref{eq:gauge23matrix}, with primed matrices given by
\begin{equation}\label{eq:MNtransf}
    A' = M^{-1} A M\, , \quad B' = M^{-1} B N\, , \quad C' = N^{-1} C M\, , 
    \quad D' = N^{-1} D N\, .
\end{equation}
Note that this transformation leaves equations \eqref{eq:matrixeqs} invariant, 
as it should. Therefore, two solutions of the matrix equations \eqref{eq:matrixeqs} 
that differ by a 
transformation of the above form must be regarded as equivalent.
Also note that, if $(A,B,C,D)$ is a solution of the system \eqref{eq:matrixeqs}, 
then so is $(-A,-B,-C,-D)\,$. 
Notice that the transformation 
$(A,B,C,D) \mapsto (A, -B, -C, D)\,$, that is a symmetry of the system 
\eqref{eq:matrixeqs}, 
can be generated by a $GL(2,\mathbb{R})\times GL(2,\mathbb{R})$ transformation
with $(M,N) = (\mathbb{I} , -\mathbb{I})\,$.

The algebraic problem at hand -- classifying matrices $A$, $B$, $C$ $D$ 
satisfying \eqref{eq:matrixeqs} up to the equivalences \eqref{eq:MNtransf} 
-- can be viewed as a quiver representation problem. 
The quiver in our case is
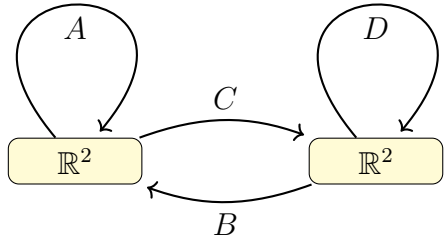
\begin{align*}
\begin{tikzpicture}
    \filldraw[black] (-4,0)  node[rectangle, rounded corners,draw,fill=yellow!20, text width=1.5cm, align=center] (A) {$\mathbb{R}^2$};
    \filldraw[black] (0,0)  node[rectangle, rounded corners,draw,fill=yellow!20, text width=1.5cm, align=center] (B) {$\mathbb{R}^2$};
    \draw[-latex,thick,-{>[sep=3pt]}] (A) to[out=20,in=160] node[anchor=south] {$C$} (B);
    \draw[-latex,thick,-{>[sep=3pt]}] (B) to[out=-160,in=-20] node[anchor=north] {$B$} (A);
    \draw[-latex,thick,-{>[sep=3pt]}] (A)  edge   [out=130,in=50,looseness=15]   node[anchor=north] {$A$} (A);
    \draw[-latex,thick,-{>[sep=3pt]}] (B)  edge   [out=130,in=50,looseness=15]   node[anchor=north] {$D$} (B);
\end{tikzpicture}
\end{align*}
and contains two vertices that each correspond to a two-dimensional space 
$\mathbb{R}^2$; the left vertex for the spin-2 sector and the right vertex for the 
spin-3 sector. The four edges correspond to the maps $A$, $B$, $C$, $D$ from a vector 
space to another. Basic definitions about quivers can be found in reference 
\cite{kirillov2016quiver} and appendix \ref{app:quivers}.

This a  wild quiver, since the underlying graph
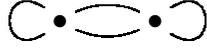
\begin{align*}
\xymatrix{
 \bullet \ar@{-}@(ul,dl)\ar@{-}@/^/[r] & \bullet \ar@{-}@(ur,dr)\ar@{-}@/^/[l]}
\end{align*}
is neither Dynkin nor Euclidean, i.e.\ does not correspond to a simply-laced 
simple Lie algebra or their affine extensions. Representations of wild quivers 
are not classified in general; however, in this particular case a full 
classification can be achieved because of the small dimensions and number of matrices 
involved. 
More generally, the problem in (A)dS$_3$ is indeed equivalent to 
identifying a finite-dimensional representation of the (anti)-de Sitter algebra
reproducing the set of fields at hand.
This is done in what follows.
The flat limit will then be studied, exhibiting a one-parameter freedom.

\paragraph{Classification of the solutions.}

Using $GL(2,\mathbb{R})\times GL(2,\mathbb{R})$ transformations
generated by the matrices $M$ and $N\,$, 
the matrices $A$ and $D$ can be put in one 
of the following three real Jordan forms:
\begin{equation}\label{eq:Jordan}
    \begin{pmatrix}
        \lambda_1 & ~0 \\
        0 & ~\lambda_2
    \end{pmatrix}\, , \quad \begin{pmatrix}
        \lambda & ~1 \\
        0 & ~\lambda
    \end{pmatrix}\quad \text{or} \quad
    \begin{pmatrix}
        \mu & ~-\nu \\ \nu & ~\mu
    \end{pmatrix}\, ,
\end{equation}
where we insist that all the entries are real. 
The last of those three forms is similar to a complex diagonal matrix 
with complex conjugate eigenvalues $\mu \pm i \nu\,$.

A detailed analysis shows that there are two general classes of 
solutions.
\begin{enumerate}
    \item The four matrices $A, B, C$ and $D$ are all real diagonal 
    (first form in \eqref{eq:Jordan}). This requires $\sigma=1\,$, the AdS$_3$ 
    background.
    \item In the second case, they are all antisymmetric (third form in 
    \eqref{eq:Jordan} with $\mu = 0$). This is possible only in the dS$_3$ 
    background, i.e. $\sigma = -1$.
\end{enumerate}
In the flat case, which formally amounts to taking $\sigma=0\,$, it is easy 
to show that we exactly recover the results of the section 
\ref{sec:firstorderreformulation}, 
formulae \eqref{eq:gaugespin3} and \eqref{eq:gaugespin2}. 

\noindent \underline{Fully diagonal case:} If the four 
matrices are diagonal, then the fields $(e^a,E^{aa})$ and $(\omega^a, \Omega^{aa})$ 
form two separate systems.\footnote{Recall that to reach the form \eqref{eq:Jordan} 
for the matrices entering the gauge transformations \eqref{eq:gauge23matrix} we 
allowed redefinitions of fields and parameters, see eq.~\eqref{eq:fieldredefMN}. 
For the spin-2 sector, for instance, the fields in each separate system can 
originate from linear combinations of the non-linear vielbein and spin-connection.} 
Writing any of those generically as $(f^a, F^{aa})$, they have gauge transformations 
of the form
\besubeqs \label{eq:transf_diag}
\begin{align}
        \delta f^a &= \nabla \epsilon^a + \lambda\, a\, \varepsilon^{abc} \,h_b 
        \,\epsilon_c + \lambda \,b \, h_b\, \epsilon^{ab}\;,\\[5pt]
    \delta F^{aa} &= \nabla \epsilon^{aa} + \lambda \,c\, 
    \left( h^a\,\delta^a_b - \tfrac{1}{3}\, \eta^{aa}\,h_b\right) 
    \epsilon^b + \lambda \,d\, \varepsilon^{abc}\, h_b \epsilon_c{}^{a} .
\end{align}
\esubeqs
Here, the parameters $a$, $b$, $c$, $d$ are real numbers (the diagonal elements 
of the corresponding matrices) constrained to satisfy
\besubeqs \label{eq:quadraticeqsabcd}
\begin{align}
    a^2 - \tfrac{5}{6}\, b c &= \sigma \, , \\
    \tfrac{1}{2}\, d^2 + c b &= 2 \sigma \, , \\
    -\tfrac{3}{2}\, d c + c a &= 0\, , \\
    -\tfrac{3}{2}\, b d + a b &= 0\, .
\end{align}
\esubeqs
These equations only admit solutions in AdS$_3$, i.e.\ for $\sigma = + 1$. 
This conclusion is reached from an analysis of the free equations of motion, but it 
anticipates the option to define interacting higher-spin gauge theories in AdS$_3$ 
from the sum of two non-Abelian Chern-Simons actions, that  stems from the structure 
of the isometry algebra of AdS$_3$. The latter is not simple, 
$so(2,2) \cong sl(2,\mathbb{R}) \oplus sl(2,\mathbb{R})$, and this allows one, 
e.g., to rewrite the Einstein Hilbert action as the difference of two 
$sl(2,\mathbb{R})$ actions \cite{Achucarro:1987vz, Witten:1988hc}. 
The two spin-2 fields belonging to the two separate systems discussed above 
are thus the analogues of the two connections entering the two 
$sl(2,\mathbb{R})$ actions of \cite{Achucarro:1987vz, Witten:1988hc} 
or the generalisations thereof studied, e.g., in \cite{Blagojevic:2003vn}.

There are solutions of \eqref{eq:quadraticeqsabcd}
where the spin $2$ and spin $3$ sectors of the 
system do not mix:
\begin{equation}\label{eq:abcdsolutionnomix}
    a = 1\, , \quad b = 0\, , \quad c = 0\, , \quad d = \pm 2\, .
\end{equation}
This case corresponds to the free limit of a $sl(3,\mathbb{R})$ action. 
Combining the two separate systems as, e.g., in \cite{Campoleoni:2010zq} 
one obtains a model that can be deformed into an interacting higher-spin 
theory described by a $sl(3,\mathbb{R}) \oplus sl(3,\mathbb{R})$ Chern-Simons 
theory.\footnote{In this context the sign freedom in \eqref{eq:abcdsolutionnomix} 
has a neat interpretation: one can indeed introduce the connections 
$f^a_\pm = \omega^a \pm \lambda e^a$ for the spin-2 sector and then choose 
to define the corresponding connections for the spin-3 sector either as 
$f^{aa}_\pm = \omega^{aa} \pm \lambda e^{aa}$ or as 
$f^{aa}_\pm = \omega^{aa} \mp \lambda e^{aa}$. 
The latter two options correspond, respectively, to $d = 2$ and $d = - 2$ 
in \eqref{eq:transf_diag}.
In this case, the spin-2 sector is associated with a $so(2,2)$ subalgebra 
of the full gauge algebra. The spin-2 fields thus precisely correspond to 
the linearization of the two $sl(2,\mathbb{R})$ connections 
of \cite{Achucarro:1987vz, Witten:1988hc, Blagojevic:2003vn}. 
These models admit a consistent truncation to Einstein's gravity, 
contrary to the possible interacting theories based on more exotic setups 
in which different spins already mix in the free theory.}

The more interesting solution from our current perspective
has both $b$ and $c$ nonvanishing 
and mixes the two sectors (spin-2 and spin-3) of the system $(f^a,F^{aa})$. 
It corresponds to 
\begin{equation}\label{eq:abcdsolutionmix}
    a = \frac{3}{2}\, , \quad b = 1\, , \quad c = \frac{3}{2}\, , \quad d = 1\, .
\end{equation}

When two systems $(f^a_{(i)},F^{aa}_{(i)})\,$, $i=1,2\,$, are considered 
simultaneously
so as to reconstruct vielbeins and spin-connections,
we can combine the solutions \eqref{eq:abcdsolutionnomix} 
or \eqref{eq:abcdsolutionmix} for each system with a relative sign. 
In the ensuing analysis we focus on the cases in which at least one of the 
two systems $(f^a_{(i)},F^{aa}_{(i)})$ mixes the sectors with different spins. 
We thus exclude the well-studied case leading to 
$sl(3,\mathbb{R}) \oplus sl(3,\mathbb{R})$ interacting theories or its 
generalisations with non-vanishing torsions \cite{Peleteiro:2020ubv} 
(both corresponding to $B = C = 0$ in \eqref{eq:gauge23matrix}, but 
involving different relative coefficients between the two sectors).
Taking into account \eqref{eq:abcdsolutionnomix}, \eqref{eq:abcdsolutionmix} 
and the relative sign introduced by the combination of the two systems, 
there exist only six inequivalent solutions  
for the matrices $A, B, C$ and $D$ in the case 
where they are all real diagonal and $B$, $C$ are different from zero. 
They are explicitly given by
\begin{equation}\label{eq:soluA1}
    A_1 = \tfrac{3}{2}\begin{pmatrix}
         1 & ~0\; \\ 0 & ~\eta \;
    \end{pmatrix}\, , \quad
    B_1 = \begin{pmatrix}
        1 & ~0 \\ 0 & ~\eta
    \end{pmatrix}\, , \quad
    C_1 = \tfrac{3}{2}\begin{pmatrix}
        1 & ~0 \\ 0 & ~\eta
    \end{pmatrix}\, , \quad
    D_1 = \begin{pmatrix}
         1 & ~0\; \\ 0 & ~\eta\;
    \end{pmatrix} \quad (\eta = \pm 1\, , \; \sigma = +1)
\end{equation}
when the two systems both mix spin $2$ and spin $3$, and
\begin{equation}\label{eq:soluA2}
    A_2 = \begin{pmatrix}
         \tfrac{3}{2} & ~0\; \\ 0 & ~\eta_1 \;
    \end{pmatrix}\, , \quad
    B_2 = \begin{pmatrix}
        1 & ~0 \\ 0 & ~0
    \end{pmatrix}\, , \quad
    C_2 = \begin{pmatrix}
        \tfrac{3}{2} & ~0 \\ 0 & ~0
    \end{pmatrix}\, , \quad
    D_2 = \begin{pmatrix}
         1 & ~0\; \\ 0 & ~2 \eta_2 \;
    \end{pmatrix} \quad (\eta_i = \pm 1\, , \; \sigma = +1) 
\end{equation}
when only one of them does, say the first one $(e^a, E^{aa})$.
We recall that these solutions only exist in AdS$_3$ space, $\sigma = 1$.

\noindent \underline{Antisymmetric case:} Apart from the fully diagonal, real cases
presented above, the only other real solution of \eqref{eq:matrixeqs} 
with $B$ and $C$ different from zero is
\begin{equation}\label{eq:soluA3}
    A_3 = \tfrac{3}{2} \begin{pmatrix}
          0 & ~1\; \\ -1 & ~0 \;
    \end{pmatrix}\, , \quad
    B_3 = \begin{pmatrix}
        0 & ~1\; \\ -1 & ~0
    \end{pmatrix}\, , \quad
    C_3 = \tfrac{3}{2} \begin{pmatrix}
          0 & ~1\; \\ -1 & ~0 \;
    \end{pmatrix}\, , \quad
    D_3 = \begin{pmatrix}
          0 & ~1\; \\ -1 & ~0 \;
    \end{pmatrix} \quad (\sigma = -1)\, . 
\end{equation}
This solution exists only in the dS background, $\sigma = -1$.
This completes the classification of the real solutions with mixing 
of the system of equations \eqref{eq:matrixeqs}.

Note that, using $GL(2,\mathbb{R})\times GL(2,\mathbb{R})$ transformations
generated by the matrices $M$ and $N\,$, 
the solution \eqref{eq:soluA1} in the case $\eta = -1$ 
and the solution \eqref{eq:soluA3} can be brought in a unified 
anti-diagonal form, which has the advantage of being valid for 
\emph{both} signs of the cosmological constant:
\begin{equation}\label{eq:simplesol23}
    A_0 = \begin{pmatrix}
        0 & ~1\; \\ \frac{9\sigma}{4} & ~0\;
    \end{pmatrix}\, , \quad
    B_0 = \begin{pmatrix}
        0 & ~1 \\ \frac{3\sigma}{2} & ~0
    \end{pmatrix}\, , \quad
    C_0 = \begin{pmatrix}
        0 & ~1 \\ \frac{3\sigma}{2} & ~0
    \end{pmatrix}\, , \quad
    D_0 = \begin{pmatrix}
        0 & ~1 \\ \sigma & ~0
    \end{pmatrix}\, . 
\end{equation}
This form does not cover the case $\eta = +1$ of solution \eqref{eq:soluA1}, 
or solution \eqref{eq:soluA2}.

\paragraph{Gauge-invariant action in (A)dS.}

We look for an action in the form
\begin{equation}\label{eq:action23ads}
    S[e^a,\omega^a,E^{aa},\Omega^{aa}] = 
    \frac{1}{2\lambda} \int_{M_3} \left[ \begin{pmatrix}
        \lambda e_a & \; & \omega_a
    \end{pmatrix} G \begin{pmatrix}
        \lambda R^a(e) \\ R^a(\omega)
    \end{pmatrix} +
    \begin{pmatrix}
        \lambda E_{aa} & \; & \Omega_{aa}
    \end{pmatrix} H \begin{pmatrix}
        \lambda R^{aa}(E) \\ R^{aa}(\Omega)
    \end{pmatrix} \right]\, .
\end{equation}
The $2\times2$ matrices $G$ and $H$ should be symmetric and non-degenerate. 
The constraints arising from gauge invariance of the action read
\besubeqs\label{eq:actionmetric23}
\begin{align}
    A^T G - G A  &= 0 \\
    GB + C^T H &= 0 \\
    D^T H - H D &= 0\, ,
\end{align}
\esubeqs
with solutions related by a transformation of the form
\begin{equation}\label{eq:MNtransfGH}
    G' = M^T G M\, , \quad H' = N^T H N
\end{equation}
being considered equivalent (since they differ by the field redefinition 
\eqref{eq:fieldredefMN}). We now discuss the solutions for $G$ and $H$ 
corresponding to the solutions for the matrices $A$, $B$, $C$, $D$ found above.

\noindent
\underline{Fully diagonal case:}
For the six solutions \eqref{eq:soluA1} 
and \eqref{eq:soluA2} in AdS$_3$, one can show that, 
by the action of residual transformations by matrices $M$ and $N\,$,  
the matrices $G$ and $H$ can be taken to be diagonal. 
Therefore, we can again look at subsystems $(f^a, F^{aa})$ in 
isolation.\footnote{While this is certainly true for the free theory, 
we stress that not all combinations may allow one to introduce non-linear 
deformations. For instance, in the case of AdS$_3$ gravity, one introduces the 
$sl(2,\mathbb{R})$-valued connections $f^a_\pm = \omega^a \pm \lambda e^a$. 
The different dependence on the vielbein has an impact on the signs entering 
the linearized gauge transformations that read 
$\delta f^a_\pm = \nabla \lambda_\pm \pm \varepsilon^{ab}{}_c h_b \lambda^c_\pm$, 
thus suggesting the need for a given relative sign to allow for a non-linear 
completion.} 
The equations \eqref{eq:actionmetric23} then reduce to the single constraint 
$gb+ch = 0$ for the diagonal coefficients (written here $g$ and $h$ generically).

When the system $(f^a, F^{aa})$ does not mix the spin $2$ and spin $3$ fields 
(solution \eqref{eq:abcdsolutionnomix}), we get the sum of two decoupled actions.  
The coefficients $g$ and $h$ can be rescaled by a field redefinition (and/or an 
overall factor in the action), leaving only a relative sign: $g = 1$, $h = \pm 1$. 
The action explicitly reads 
\begin{equation}\label{eq:actionfFnomix}
    S[f^a,F^{aa}] = 
    \frac{1}{2} \int_{M_3} \left( f_a R^a(f) \pm F_{aa} R^{aa}(F) \right)\, ,
\end{equation}
with gauge invariance and curvatures
\besubeqs
\begin{align}
        \delta f^a &= \nabla \epsilon^a + \lambda\, \varepsilon^{abc} \,h_b \,\epsilon_c \;,&\quad R^a(f) &= \nabla f^a + \lambda\, \varepsilon^{abc} \,h_b \,f_c \;,\\
    \delta F^{aa} &= \nabla \epsilon^{aa} \pm 2 \lambda \, \varepsilon^{abc}\, h_b \epsilon_c{}^{a}\;,&\quad R^{aa}(F) &= \nabla F^{aa} \pm 2 \lambda \, \varepsilon^{abc}\, h_b F_c{}^{a}\;.
\end{align}
\esubeqs

When the system $(f^a, F^{aa})$ mixes the spin $2$ and spin $3$ sectors 
(solution \eqref{eq:abcdsolutionmix}), one finds $g = 1$ and $h = -\tfrac{2}{3}\,$. 
The action is then
\begin{equation}
    S[f^a,F^{aa}] = 
    \frac{1}{2} \int_{M_3} \left( f_a R^a(f) - \tfrac{2}{3} F_{aa} R^{aa}(F) \right)\, ,
\end{equation}
with gauge invariance and curvatures
\besubeqs
\begin{align}
        \delta f^a &= \nabla \epsilon^a + \tfrac{3}{2}\lambda\, \varepsilon^{abc} \,h_b \,\epsilon_c + \lambda \, h_b\, \epsilon^{ab} \;,\\[5pt]
    \delta F^{aa} &= \nabla \epsilon^{aa} + \tfrac{3}{2}\lambda\, 
    \left( h^a\,\delta^a_b - \tfrac{1}{3}\, \eta^{aa}\,h_b\right) 
    \epsilon^b + \lambda \, \varepsilon^{abc}\, h_b \epsilon_c{}^{a}\;,\\[5pt]
    R^a(f) &= \nabla f^a + \tfrac{3}{2}\lambda\, \varepsilon^{abc} \,h_b \,f_c + \lambda \, h_b\, F^{ab} \;,\\[5pt]
    R^{aa}(F) &= \nabla F^{aa} + \tfrac{3}{2}\lambda\, 
    \left( h^a\,\delta^a_b - \tfrac{1}{3}\, \eta^{aa}\,h_b\right) 
    f^b + \lambda \, \varepsilon^{abc}\, h_b F_c{}^{a}\;.
\end{align}
\esubeqs
We can now put two such systems together, corresponding to $(e^a, E^{aa})$ and 
$(\omega^a, \Omega^{aa})$, possibly with relative signs: the matrices $G$ and 
$H$ are then
\begin{equation}\label{eq:GH1}
    G_1 = \begin{pmatrix}
        1 & ~0 \\ 0 & ~\tau
    \end{pmatrix}\, , \quad H_1 = - \frac{2}{3}\begin{pmatrix}
        1 & ~0 \\ 0 & ~\tau
    \end{pmatrix} \end{equation}
for \eqref{eq:soluA1} and
\begin{equation}\label{eq:GH2}
    G_2 = \begin{pmatrix}
        1 & ~0 \\ 0 & ~\tau_1
    \end{pmatrix}\, , \quad H_2 = \begin{pmatrix}
        - \tfrac{2}{3} & ~0 \\ 0 & ~\tau_2
    \end{pmatrix}  
\end{equation}
for solution \eqref{eq:soluA2}, with independent signs $\tau$, 
$\tau_1$, $\tau_2 = \pm 1\,$. 
This whole discussion is valid in AdS$_3$ space only ($\sigma = +1$).

\noindent
\underline{Antisymmetric case:}
The remaining case is that of solution \eqref{eq:soluA3}, 
which is valid in dS$_3$ space only ($\sigma = -1$). 
As explained above, using a transformation generated by appropriate matrices 
$M$ and $N$, this case is covered by the solution \eqref{eq:simplesol23} 
for $\sigma = -1$. Therefore, the matrices $G_3$ and $H_3$ corresponding to the 
solution \eqref{eq:soluA3} are not presented explicitly since they can be obtained 
from the matrices $G_0$ and $H_0$ written in \eqref{G0H0} below.

\noindent
\underline{Off-diagonal case:}
We now consider the solution 
\eqref{eq:simplesol23}, which is valid in both dS$_3$ and AdS$_3$ and covers the case 
\eqref{eq:soluA1} with $\eta = -1$ (in AdS$_3$) as well as the 
case \eqref{eq:soluA3} (in dS$_3$). 
With the matrices $A_0$, $B_0$, $C_0$ and $D_0$ of \eqref{eq:simplesol23} 
considered above, the solution of equations \eqref{eq:actionmetric23} is given by
\begin{equation}
\label{G0H0}
    G_0 = \begin{pmatrix}
        0 & ~1 \\ 1 & ~0
    \end{pmatrix} = - H_0\, ,
\end{equation}
up to the action of matrices $M$ and $N$ that leave \eqref{eq:simplesol23} invariant. 
These matrices $G_0$ and $H_0$ reproduce the standard symplectic structures 
$e \,{\rm d} \omega$ and $E \,{\rm d}\Omega\,$, respectively.

To conclude, we have found seven inequivalent systems mixing the spin-2 
and spin-3 gauge transformations in (A)dS$_3$ backgrounds. 
Of those seven solutions, two can be unified in a form 
\eqref{eq:simplesol23} -- \eqref{G0H0} that is valid for both signs of the 
cosmological constant. The other five exist in AdS$_3$ space only. 
A representation-theoretic argument for the existence of a discrete family 
of systems will be explained in section \ref{sec:general}.

In what follows we first consider the flat limit of the field equations, 
and then, the flat limit of the corresponding actions.

\paragraph{Flat limit I: curvatures and gauge transformations.}

To recover the flat limit \eqref{eq:gaugespin3} -- \eqref{eq:gaugespin2}
of the curvatures and gauge transformations, 
some parameters are fixed as functions of $x$ and~$\gamma\,$:
\begin{equation}\label{eq:flatlimit}
    x_2 = -1\, , \quad x_4 = 2z\, , \quad x_{10} = -3x 
    = \frac{2 \gamma  z}{3 \left(3 \gamma  z^2-2\right)}\, , \quad x_{12} 
    = \frac{4}{3}\, .
\end{equation}
At first sight, it appears that these values for the above four 
parameters do not comply with any of the solutions presented above. 
However, the values \eqref{eq:flatlimit} can be reached by acting 
on the simple solution \eqref{eq:simplesol23} with a 
$GL(2,\mathbb{R})\times GL(2,\mathbb{R})$ transformation 
of the form \eqref{eq:MNtransf} 
with $z$-dependent matrices $M$ and $N$ given by
\begin{equation}\label{eq:matricesMandN}
    M = \begin{pmatrix}
        \frac{3}{\sqrt{2}}\,\Delta & \; & z\vspace{0.1cm} \\ -\frac{9\sigma}{4}\,z & 
        \; & -\frac{3}{\sqrt{2}}\,\Delta
    \end{pmatrix} \, , \quad N = \begin{pmatrix}
        0 & \; & 1 \\ \frac{3\sigma}{4} & \; & 0
    \end{pmatrix}\, ,
\end{equation}
where $\Delta$ is the square root
\begin{equation}
    \Delta = \sqrt{\gamma \sigma \left(2 \gamma  z^2-1\right)}\, .
\end{equation}

Note that we should be looking at \emph{real} solutions for 
the parameters $x_i\,$. 
The reality of $\Delta$ then determines whether the  
field equations \eqref{eq:eomflat} around Minkowski space
can be extended to dS$_3$ ($\sigma = -1$) and/or AdS$_3$ ($\sigma = 1$):
\begin{itemize}
    \item If $\gamma = +1$, we have 
    $\Delta = \sqrt{ \sigma  z^2 \left(2 z^2-1\right)}\,$: the model 
    can be extended to dS$_3$ when $z^2< 1/2\,$, 
    to AdS$_3$ for $z^2 >1/2\,$, and to both when $z^2 = 1/2\,$. 
    In particular, the original action of \cite{Boulanger:2020yib} 
    corresponds to $z = -1$ and therefore 
    can only be continued to AdS$_3\,$, not to dS$_3\,$. 
    \item If $\gamma = -1\,$, we have 
    $\Delta = \sqrt{ \sigma  z^2 \left(2 z^2 + 1\right)}\,$: 
    these models can only be deformed to AdS$_3\,$.
\end{itemize}
Equivalently, this means that the field equations \eqref{eq:eomflat} 
obtained from the one-parameter family of actions 
\eqref{eq:actionfirstordercurvatures} 
can be reached from the simple solution \eqref{eq:simplesol23} 
in dS$_3$ or AdS$_3$ (depending on the values of $\gamma$ and 
$z$ discussed above) by \emph{first} performing a 
$GL(2,\mathbb{R})\times GL(2,\mathbb{R})$
field redefinition of the model in curved space, 
\emph{and then} taking the flat limit.

To conclude this discussion, let us also cover the isolated case 
\eqref{eq:a00} with $a_0 = 0$, which corresponds to 
\begin{equation}\label{eq:flatlimit0}
    x_2 = -1\, , \quad x_4 = 2z =  \sqrt{2}\, , \quad x_{10} = -3x = 
    \frac{2\sqrt{2}}{9}\, , \quad x_{12} = \frac{4}{3}\, .
\end{equation}
This model can only be continued to AdS ($\sigma = +1$). 
To reach the values \eqref{eq:flatlimit0} from the simple solution 
\eqref{eq:simplesol23}, the matrices $M$ and $N$ can be taken as
\begin{equation}
    M = \begin{pmatrix}
        0 & \; & - \frac{3}{\sqrt{2}} \\ \frac{27}{4\sqrt{2}} & \; & 0
    \end{pmatrix} \, , \quad N = \begin{pmatrix}
        -2 & \; & 1 \\ \frac{3}{4} & \; & - \frac{3}{2}
    \end{pmatrix}\, .
\end{equation}

\paragraph{Flat limit II: action.}

The previous discussion only applies to the gauge transformations 
and curvatures, i.e.~at the level of equations of motion. 
As we shall see, the existence of an action is much more constrained.

In the action \eqref{eq:action23ads}, we have adjusted the powers of $\lambda$ such 
that the terms appearing in \eqref{eq:actionfirstordercurvatures} come with 
$\lambda^0$. Then, the terms $e_a R^a(e)$ and $E_{aa} R^{aa}(E)$ come with 
$\lambda^{1}$ and vanish as $\lambda \to 0$, while the terms $\omega_a R^a(\omega)$ 
and $\Omega_{aa} R^{aa}(\Omega)$ come with $\lambda^{-1}$ and are singular in the flat 
limit.  Therefore, the action \eqref{eq:action23ads} has a smooth flat limit if the 
bottom-right entry of the matrices $G$ and $H$ vanishes: 
$G_{22} = 0 = H_{22}\,$.

To recover the action \eqref{eq:actionfirstordercurvatures} in the flat limit, 
we should therefore impose
\begin{equation}\label{eq:flatlimitaction}
    G_{22} = 0 = H_{22} \, , \quad G_{12}= 1 = G_{21} \, , 
    \quad H_{12} = \frac{2z}{3x}=H_{21}\;,
\end{equation}
in addition to the conditions \eqref{eq:flatlimit} (one can also impose the 
opposite of the values in \eqref{eq:flatlimitaction}, since the global sign 
of the action is of no relevance here). 
Remarkably, the system then only admits solutions for specific values of the 
product $xz\,$:
\begin{equation}
    x z = -\frac{2}{3}\, , \; -\frac{2}{15}\, , \; \frac{2}{45}\, , \; \frac{2}{9}\, .
\end{equation}
Of those values, only $xz = 2/9$ is possible in both dS$_3$ ($\sigma = -1$) 
and AdS$_3$ ($\sigma = 1$) spaces. 
Recalling that $x = -\tfrac{2 \gamma  z}{9 \left(3 \gamma  z^2-2\right)}\,$, 
the equality $xz = \frac{2}{9}$ implies that $\gamma\,z^2 = \frac{1}{2}\,$, 
which in turn implies that $\gamma = +1$ and $z=\pm \frac{1}{\sqrt{2}}\,$.
The other values of $xz$ correspond to solutions in AdS$_3$ space only. 
The dual system of \cite{Boulanger:2020yib} has $xz = -2/9$, hence 
cannot be deformed to (A)dS$_3$.

These solutions can most efficiently be described by exhibiting 
the matrices $M$ and $N$ that can be used to reach them from some 
elementary solution presented above.
\begin{itemize}
    \item  In both dS$_3$ and AdS$_3$, the solution with $x z = 2/9$ 
    can be reached from the simple solution \eqref{eq:simplesol23}-- \eqref{G0H0} 
    by acting with the matrices
    \begin{equation}
        M = \begin{pmatrix}
        -1 & \; & 0 \\ 0 & \; & 1
    \end{pmatrix} \, , \quad N = -z\begin{pmatrix}
        \frac{3}{2} & \; & 0 \\ 0 & \; & 2
    \end{pmatrix}\, .
    \end{equation}
    \item The other values are in the orbit of the rather strange solution 
    \eqref{eq:soluA2}--\eqref{eq:GH2}, with signs $\tau_1 = -1$ and $\tau_2 = +1$ 
    for the matrices $G$ and $H$. The different choices of signs $\eta_1$ and $\eta_2$ 
    provide the different values for the product $xz\,$:
    \begin{equation}
        xz = - \frac{2}{3 (3-2 \eta_1) (2 \eta_2-1)} 
        \in \{-\frac{2}{3}\, , \; -\frac{2}{15}\, , \; \frac{2}{45}\, , 
        \; \frac{2}{9}\}\;,
    \end{equation}
    (in particular, $\eta_1 = - \eta_2 = 1$ provides another inequivalent solution 
    with $xz = 2/9$). The matrices $M$ and $N$ are
    \begin{equation}
        M = \begin{pmatrix}
        -\sqrt{\frac{3-2\eta_1}{2}} & \; & \sqrt{\frac{2}{3-2\eta_1}} \\ 0 & \; & - 
        \sqrt{\frac{2}{3-2\eta_1}}
    \end{pmatrix} \, , \quad N = z\sqrt{3-2\eta_1} \begin{pmatrix}
        \frac{3(2\eta_2-1)}{2\sqrt{2}} & \; & - \sqrt{2} \\ 0 & \; & 
        -\frac{2}{\sqrt{3}}
    \end{pmatrix}\, .
    \end{equation}
\end{itemize}

We therefore conclude that there is only a discrete set of values of 
the free parameter $z$ such that the action \eqref{eq:actionfirstordercurvatures} 
in Minkowski spacetime admits a deformation to (A)dS$_3\,$. 
Of those values, only $\gamma = +1$ and $z=\pm \frac{1}{\sqrt{2}}\,$ 
can be deformed to both dS and AdS; the resulting actions are in the orbit 
of the system described by the matrices \eqref{eq:simplesol23}--\eqref{G0H0}.

\section{Chern-Simons formulation and Generalizations}
\label{sec:general}
In sections \ref{sec:propagating} and \ref{sec:topological} a number 
of peculiar theories with and without propagating degrees of freedom was 
discussed, some of which were given a Chern-Simons-like formulation in section 
\ref{sec:topological}. According to \cite{Grigoriev:2020lzu} all three-dimensional 
higher spin theories 
without propagating degrees of freedom (called topological here) are equivalent to 
Chern-Simons theories. The goal of the present section is to 
develop a formalism to construct topological higher spin models 
in order to explain the examples of section \ref{sec:topological} and to find 
generalizations thereof.

\subsection{Higher spin Quivers}
We would like to describe the space of all topological higher spin theories in 
$3D$.\footnote{Some of the results below may apply to non-topological systems, e.g., 
topologically massive gravity (contrary to its name, it is not topological in the 
sense of having propagating degrees of freedom) and its various generalizations and 
higher spin extensions. See \cite{Boulanger:2014vya} for more detail on these cases.} 
We will consider only manifestly Lorentz covariant theories and, therefore, assume 
that all fields can be decomposed into a number of Lorentz (spin)-tensors. This means 
that we will consider fields that carry various representations of $so(1,2)\sim sl_2$, 
i.e. we will have a set of fields
\begin{align}
    \Phi^{\ga_1\cdots\ga_N|\aI}(x)\equiv \Phi^{\ga(N)|\aI}(x)
\end{align}
that are symmetric spin-tensors with indices $\alpha_1,\ldots,\alpha_N$, which may 
carry some additional 
label $\aI$ --- whose range of values may depend on $N$ ---  
to be able to distinguish different fields valued 
in the same $sl_2$-module. The fields can be $p$-forms with $p=0,1,2,3$. The fermions 
correspond to odd $N$ and are Grassmann odd, which is irrelevant for the free 
equations.

\paragraph{Relevant algebras.} Lorentz symmetry, i.e. $sl_2(\mathbb{R})$, 
is always manifest in our approach. Massless fields fall into representations of 
a larger algebra: the Poincar\'e algebra $iso(2,1)$ in flat space, 
$sl_2(\mathbb{C})\cong so(3,1)$ in de Sitter space and 
$sl(2,\mathbb{R})\oplus sl(2,\mathbb{R})\cong so(2,2)$ in anti-de Sitter space. 
Let $(j_1,j_2)$ denote the $(2j_1+1)(2j_2+1)$-dimensional representation of the 
(anti)-de Sitter algebra. (Partially)-Massless fields in (anti)-de Sitter space 
require representations that are 
nontrivially charged under the diagonal Lorentz algebra $sl(2,\mathbb{R})$ contained 
in the (anti)-de Sitter algebra  
\cite{Blencowe:1988gj, Bergshoeff:1989ns, Campoleoni:2010zq, Henneaux:2010xg}. 
In order to cover partially-massless fields 
\cite{Deser:1983mm,Higuchi:1986wu,Deser:2001us,Skvortsov:2006at,Buchbinder:2012bz} 
one needs \cite{Grigoriev:2020lzu} to include representations of type 
$(N,M)$, $MN\neq0$ charged under both algebras, 
i.e., to consider fields $\Phi^{\ga_1...\ga_N,\gad_1...\gad_M|\aI}(x)$. 
(Un)dotted indices are those of $sl(2,\mathbb{R})\oplus sl(2,\mathbb{R})$ or of 
$sl_2(\mathbb{C})$. The massless case corresponds to $(M,0)$, $(0,M)$. To make the 
link to the metric-like formulation of (partially)-massless fields one has to consider 
conjugated pairs $(M,N)\oplus(N,M)$. In what follows, we only keep the Lorentz 
symmetry manifest. In particular, this is the only option for the Poincar\'e case.

Finite-dimensional representations of all algebras mentioned above save for $iso(2,1)$ 
are completely reducible. It is the Poincare case that is tricky and admits a lot of 
strange topological higher spin systems. It is worth adding that a classification of 
finite-dimensional representations of $iso(2,1)$ is not available.
Some examples and references can be found in section~3 of \cite{Campoleoni:2021blr}.

\paragraph{General topological systems.} It is useful to pack $\Phi^{\ga(N)|\aI}(x)$ 
into a generating function by contracting the $\ga$'s with an auxiliary spinor $y_\ga$:
\begin{align}
    \Phi(y|x)&= \sum_{N,{\cal I}} \tfrac{1}{N!}\, y_{\ga_1} \cdots 
    y_{\ga_N}\,\Phi^{\ga_1\cdots \ga_N|{\cal I}}(x)\,.
\end{align}
Different representations of $sl_2$ belong to the eigenspaces of the Euler 
operator $N=y^\alpha\frac{\partial}{\partial y^\alpha}\,$. 
To allow for multiplicity, accounted by the index $\aI$, 
we assume that for each eigenvalue of $N$, the field $\Phi(y)$ takes values 
in some vector space $V_N$.

Let us assume that $\mathcal{M}_3$ is a three-dimensional space equipped with 
a dreibein $h^{\ga\gb}$ and a compatible spin-connection and, hence, we have a 
Lorentz covariant derivative $\nabla$. The most general topological system we 
can write for $\Phi$s all having the same form degree is:\footnote{Similar systems 
of varying degree of generality have already been considered in the literature 
\cite{Vasiliev:1992gr, Boulanger:2014vya}. }
\begin{align}
    \nabla \Phi&= Q\Phi\,,
\end{align}
where the most general horizontal differential $Q$ reads
\begin{align}\label{eq:Q}
    Q\Phi&= [\alpha(N) h^{\ga\ga} y_\ga y_\ga +\beta(N) h^{\ga\ga} \pl_\ga \pl_\ga 
    +\gamma(N) h^{\ga\ga} y_\ga \pl_\ga]\Phi\,,
\end{align}
where $\partial_\alpha \equiv \frac{\partial}{\partial y^\alpha}$.
Here $\alpha_N,\beta_N,\gamma_N$ are linear maps 
\begin{align}
    \alpha_N: \ V_{N-2}\rightarrow V_N\,, \qquad
    \beta_N: \ V_{N+2}\rightarrow V_N\,, \qquad
    \gamma_N: \ V_{N}\rightarrow V_N\,,
\end{align} 
i.e.\ they are matrices that depend (including the size) on 
$N=y^\alpha \pl_\alpha$. For the system to be topological, the covariant derivative 
$D=\nabla-Q$ has to be nilpotent. The same condition implies it is gauge invariant 
under $\delta \omega=D\xi$, where $\xi$ are zero-forms taking values in the same 
collection of vector spaces $V_N$. The nilpotency condition gives a number of 
conditions:
\besubeqs\label{integrability}
\begin{align}
    (N-2)\alpha(N) \gamma(N-2) = (N+2)\gamma(N) \alpha(N)\,,\\[5pt]
    (N-2)\gamma(N-2)\beta(N-2)  = (N+2)\beta(N-2)\gamma(N)\,, \\[5pt]
    -(N-1) \alpha(N)\beta(N-2) +\gamma(N)\gamma(N) +(N+3)\beta(N)\alpha(N+2)  
    = \sigma \lambda^2 \mathbf{1}\,.\label{lastcond}
\end{align}
\esubeqs
We assumed that $\nabla^2=\sigma\,\lambda^2 H^{\ga\ga}y_\ga \pl_\ga \mathbf{1}_N$, 
where $\lambda^2$ is the cosmological constant and $\mathbf{1}_N$ is the identity 
map on $V_N$. The system has natural automorphisms that originate from linear field 
redefinitions $\Phi\rightarrow A_N\Phi$, where $A_N: V_N\rightarrow V_N$ is an 
automorphism of $V_N$:
\begin{align}
    \alpha_N&\rightarrow A^{-1}_N\alpha_N A_{N-2}\,,&
    \beta_N&\rightarrow A^{-1}_N\beta_N A_{N+2}\,,&
    \gamma_N&\rightarrow A^{-1}_N\gamma_N A_N\,.
    \label{eq:4.7}
\end{align}
Therefore, the system corresponds to a quiver with certain additional 
restrictions given by \eqref{integrability}. The quiver is
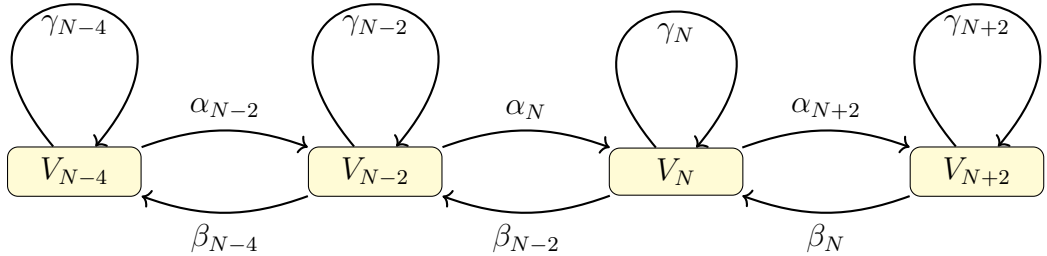
\begin{align*}
\begin{tikzpicture}
    \filldraw[black] (-4,0)  node[rectangle, rounded corners,draw,fill=yellow!20, text width=1.5cm, align=center] (A) {$V_{N-2}$};
    \filldraw[black] (0,0)  node[rectangle, rounded corners,draw,fill=yellow!20, text width=1.5cm, align=center] (B) {$V_N$};
    \filldraw[black] (4,0)  node[rectangle, rounded corners,draw,fill=yellow!20, text width=1.5cm, align=center] (C) {$V_{N+2}$};
    \draw[-latex,thick,->] (B) to[out=-160,in=-20] node[anchor=north] {$\beta_{N-2}$} (A);
    \draw[-latex,thick,->] (A) to[out=20,in=160] node[anchor=south] {$\alpha_N$} (B);
    \draw[-latex,thick,->] (B) to[out=20,in=160] node[anchor=south] {$\alpha_{N+2}$} (C);
    \draw[-latex,thick,->] (C) to[out=-160,in=-20] node[anchor=north] {$\beta_N$} (B);
    \draw[-latex,thick,->] (B)  edge   [out=130,in=50,looseness=15]   node[anchor=north] {$\gamma_N$} (B);
    \draw[-latex,thick,->] (A)  edge   [out=130,in=50,looseness=15]   node[anchor=north] {$\gamma_{N-2}$} (A);
    \draw[-latex,thick,->] (C)  edge   [out=130,in=50,looseness=15]   node[anchor=north] {$\gamma_{N+2}$} (C);
    \filldraw[black] (-8,0)  node[rectangle, rounded corners,draw,fill=yellow!20, text width=1.5cm, align=center] (D) {$V_{N-4}$};
    \draw[-latex,thick,->] (A) to[out=-160,in=-20] node[anchor=north] {$\beta_{N-4}$} (D);
    \draw[-latex,thick,->] (D) to[out=20,in=160] node[anchor=south] {$\alpha_{N-2}$} (A);
    \draw[-latex,thick,->] (D)  edge   [out=130,in=50,looseness=15]   node[anchor=north] {$\gamma_{N-4}$} (D);
\end{tikzpicture}
\end{align*}
which should be extended down to the minimal value of $N$ and up to the maximal, 
possibly infinite, value of $N$. 

In particular, the case studied in the previous section \ref{sec:topological} 
corresponds 
to two sectors with spin 2 and 3, respectively, hence with vector spaces 
$V_2$ and $V_4$ of dimension two each, for the two one-form fields 
in each sector. 
Associated with these two vector spaces, we therefore have the 
matrices $\alpha_4\,$, $\beta_2\,$, $\gamma_2$ and $\gamma_4\,$.
These four $2\times 2$ matrices correspond to the matrices $A\,$, $B\,$, $C\,$ and 
$D\,$ of section \ref{sec:topological}. 
Finally, equations \eqref{integrability} correspond to the conditions 
\eqref{eq:matrixeqs} found in that section, and the matrices $A_2$ and $A_4$
appearing in \eqref{eq:4.7} correspond to the matrices $M$ and $N$ of 
\eqref{eq:MNtransf}.

It is useful to rescale the maps as
\begin{align}
    \alpha_N&=\frac{\bar\alpha_N}{N-1}\,,  & \beta_N&=\frac{\bar\beta_N}{N+3}\,, & 
    \gamma_N&=\frac{\bar\gamma_N}{N(N+2)}\,,
\end{align}
and define $\bar\alpha$, $\bar\beta$ and $\bar\gamma$ as the maps that act on the 
corresponding $V_N$. Relations \eqref{integrability} can be summarized as 
\begin{align}\label{integrabilityB}
    [\bar\gamma,\bar\alpha]&=0\,, & [\bar\gamma,\bar\beta]&=0\,, &
    -\tfrac1{N+1} [\bar\alpha,\bar\beta] +\tfrac{1}{N^2(N+2)^2} \bar\gamma^2  &= 
    \sigma \lambda^2 \mathbf{1}\,.
\end{align}
In general the topological system looks intractable --- it corresponds to a quiver 
of the wild type. In practice this means that we cannot just diagonalize or Jordanize 
the matrices with the help of automorphisms as there are too few of them 
and the reduced form does not have any reasonable classification. However, the quiver 
is supplemented with Eqs. \eqref{integrabilityB}. Altogether, they imply that the 
total space of $\Phi^{\ga(N)|\aI}(x)$ forms a representation of the space-time 
symmetry algebra. In the (anti)-de Sitter case all finite-dimensional representations 
are completely reducible and, hence, the wildness of the quiver plays no role. It is 
the Poincare case that presents a problem. To illustrate the formalism let us consider 
some well-known examples.

\paragraph{Example: same spin.} For a set of same spin fields we have 
$\alpha=\beta=0$ and the quiver is 
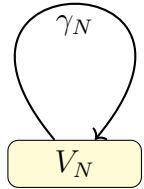
\begin{align}
\begin{tikzpicture}
    \filldraw[black] (0,0)  node[rectangle, rounded corners,draw,fill=yellow!20, text width=1.5cm, align=center] (B) {$V_N$};
    \draw[-latex,thick,->] (B)  edge   [out=130,in=50,looseness=15]   node[anchor=north] {$\gamma_N$} (B);
\end{tikzpicture}
\end{align}
It corresponds to a matrix $\gamma_N$ up to conjugation, 
$\gamma_N\rightarrow A^{-1}_N\gamma_N A_N$ and the classification is well-known: 
indecomposable representations are given by Jordan blocks (or real Jordan blocks in 
the real case). Jordan cells of size greater than two are not nilpotent and, hence, do 
not satisfy \eqref{lastcond} for any $\lambda$.

As is well-known \cite{Blencowe:1988gj, Bergshoeff:1989ns} 
and as we already recalled in section~\ref{sec:topological}, 
a single massless field in (A)dS$_3$ 
or flat space can be described (in the sense of being equivalent to the Fronsdal 
approach) by two one-forms taking values in some finite-dimensional irreducible 
representation of the $sl_2$ Lorentz algebra. Therefore $\dim V_N=2$ and $N=2s-2$. 
The matrix $\gamma_N$ can be chosen as
\begin{align}
    \textrm{(A)dS}_3&:\ \gamma_N=\begin{pmatrix} 0 & 1 \\ \sigma\lambda^2 & 
    0\end{pmatrix} , &
    \text{Minkowski}&:\ \gamma_N=F\equiv \begin{pmatrix} 0 & 1 \\ 0 & 0\end{pmatrix} .
\end{align}
Accordingly, a spin-$s$ massless field in flat space is described by
\begin{align} \label{eq:example-flat}
    d e^{\ga(2s-2)}&= h\fud{\ga}{\gb}\wedge \omega^{\gb\ga(2s-3)}\,, & 
    d\omega^{\ga(2s-2)}&=0\,,
\end{align}
where $\gamma_N=F$ is manifested by the way the fields mix with each other. It is 
invariant under 
\begin{align}\label{gaugetransmassless}
    \delta e^{\ga(2s-2)}&= d\xi^{\ga(2s-2)}-h\fud{\ga}{\gb}\wedge 
    \eta^{\gb\ga(2s-3)}\,, & 
    \delta\omega^{\ga(2s-2)}&=d\eta^{\ga(2s-2)}\,.
\end{align}
In the (A)dS$_3$ case one gets an additional term in the r.h.s.\ of the second 
equation in \eqref{eq:example-flat}:
\begin{align}
    d e^{\ga(2s-2)}&= h\fud{\ga}{\gb}\wedge \omega^{\gb\ga(2s-3)}\,, & 
    d\omega^{\ga(2s-2)}&=\sigma\lambda^2 \,h\fud{\ga}{\gb}\wedge e^{\gb\ga(2s-3)}\,.
\end{align}
As we discussed extensively in section~\ref{sec:topological}, 
in the anti-de Sitter case the system can be diagonalized by mapping it to
\begin{align}
    d A^{\ga(2s-2)}&= +\lambda h\fud{\ga}{\gb}\wedge A^{\gb\ga(2s-3)}\,, & 
    dB^{\ga(2s-2)}&=-\lambda h\fud{\ga}{\gb}\wedge B^{\gb\ga(2s-3)}\,.
\end{align}

\paragraph{Example: diagonalizable case.} Let us assume that we managed to 
diagonalize all $\alpha$, $\beta$, $\gamma$ simultaneously or, at least, 
various matrix products give the same matrix for each equation so that we can 
check the overall coefficients only. Now, the system reduces to a simple scalar 
equation. 
We assume that the module consists of $sl(2,\mathbb{R})$-tensors 
with ranks from $n_1$ to $n_2$ in steps of two:
\begin{align}
    T_{\ga(n_1)}\,, T_{\ga(n_1+2)} \,, \ldots \,, T_{\ga(n_2)}\,.
\end{align}
In this case the general solution reads \cite{Vasiliev:1992gr, Boulanger:2014vya}:
\begin{align}\label{pmsolution}
    \sigma(n)&=\frac{-\sigma\lambda^2 \left(n^2-n_1^2\right) 
    \left((n_2+2)^2-n^2\right)}{ 4n^2 \left(n^2-1\right)}\,,\\
    \gamma(n)&=\frac{\gamma_0 \lambda n_1(n_2+2)}{n (n+2)}\,, && \gamma_0=\pm 1\,,
\end{align}
where $\sigma(N)=\alpha(N)\beta(N-2)$, which is the combination invariant 
under rescalings of the fields. 

\paragraph{Example: partially-massless fields.} 
The simplest example for which the solution above is relevant is the case of 
partially-massless fields \cite{Skvortsov:2006at,Buchbinder:2012bz}, where we 
can choose 
\begin{align}\label{canonicalPM}
    \alpha_n&= \sigma(n) \, \mathrm{Id}_2\,,  &\beta_n&=\mathrm{Id}_2\,, & 
    \gamma_n&=\gamma(n) \begin{pmatrix} 0 & 1 \\ 1 & 0\end{pmatrix} \,.
\end{align}
The partially-massless system written in the usual basis reads schematically 
\begin{align}
\begin{aligned}
    \nabla \omega^k&= \gamma e^k + \alpha \omega^{k-2}+ \beta \omega^{k+2}\,,\\
    \nabla e^k&= \gamma \omega^k + \alpha e^{k-2}+ \beta e^{k+2}\,,
\end{aligned}
\end{align}
where we indicated without the explicit coefficients the contributions from the 
$\alpha$, $\beta$, $\gamma$ matrices and the number of $sl(2,\mathbb{R})$-indices 
that the fields carry. With the solution \eqref{pmsolution}, the maximal spin 
is $2s-2=n_2$ and the minimal one is $2(s-t)=n_1$, where $t$ is the depth 
of partially--masslessness. While $\alpha$ and $\beta$ are already diagonal, 
one can also diagonalize $\gamma$ to get two decoupled systems. 

In the (A)dS$_3$ case due to the complete reducibility we can map a 
partially--massless system that contains spins from $s-t$ to $s$ to two irreducible 
connections, 
$\omega^{\ga(2s-t-1),\gad(t-1)}$ and $\omega^{\ga(t-1),\gad(2s-t-1)}$ of 
$sl(2,\mathbb{R})\oplus sl(2,\mathbb{R})$. 
Therefore, any topological system in (A)dS$_3$ consists of 
(partially)-massless fields and nothing else. 

\paragraph{Free actions.} Given the equations of motion of a topological system, 
we can also ask whether they can be derived from an action principle.
Let us define a gauge-invariant curvature as $R=(\nabla -Q)\omega$ and take 
\begin{align}\label{generalaction}
    S&=\int  \langle \omega^{\aI} |G_{\aI \aJ}(N)|R^{\aJ}\rangle\,,
\end{align}
where the conjugate is defined by 
$y_\ga \rightarrow \pl_\ga$, $\pl_\ga \rightarrow -y_\ga$ (it swaps the order 
of $y$ and $\pl$) and, possibly, by complex conjugation as well. 
The gauge invariance can be checked via
\begin{align*}
    \delta S&= \langle \nabla\xi^{\aI}-Q\fud{\aI}{\aK}\xi^{\aK} |G_{\aI \aJ}(N)|R^{\aJ}\rangle=\\
    &=-\langle \xi^{\aI} |G_{\aI \aJ}(N)|QR^{\aJ}\rangle-\langle \xi^{\aK} |Q^\dag{}\fud{\aI}{\aK} G_{\aI \aJ}(N)|R^{\aJ}\rangle\,,
\end{align*}
where we integrated by parts and used the Bianchi identities $\nabla R\equiv Q R$. The gauge invariance imposes:
\besubeqs\label{eq:actionmetricgeneral}
\begin{align}
    G_{\aK \aI}(N) \beta\fud{\aI}{\aJ}(N) +G_{\aI\aJ}(N+2) \alpha\fud{\aI}{\aK}(N+2)&=0\,,\\
    G_{\aK \aI}(N) \gamma\fud{\aI}{\aJ}(N) -G_{\aI\aJ}(N) \gamma\fud{\aI}{\aK}(N)&=0\,.
\end{align}
\esubeqs
In the special case studied in Section \ref{sec:topological}, 
the corresponding equations are \eqref{eq:actionmetric23}.

These conditions imply that $Q$ is self-adjoint with respect to the bilinear 
product defined by \eqref{generalaction}. We require $G$ be symmetric, 
$G_{\aI\aJ}=G_{\aJ\aI}$, and non-degenerate for the variation to reduce to
\begin{align}
    \delta S&= 2\langle \delta\omega^{\aI} |G_{\aI \aJ}(N)|R^{\aJ}\rangle\,.
\end{align}
The equations of motion are equivalent to the desired $R^{\aI}=0$. For the 
diagonalizable case, $\sigma_N$ is a scale invariant combination of $\alpha_N$ and 
$\beta_{N-2}$, but the action principle requirement fixes the relative normalization 
for $\alpha$ and $\beta$:
\begin{align}\label{diagcase}
    (\beta_N)^2&= - \frac{\sigma_{N+2}G_{N+2}}{G_N}\,.
\end{align}
One can choose $G_N=1$, which makes the action the simplest. 
In particular, this gives an action for partially-massless fields in 
$3D$ \cite{Buchbinder:2012bz}, which is not a simple adaptation of 
\cite{Skvortsov:2006at}.

\paragraph{Example: massless fields.} For a single massless field in Minkowski space 
we can take
\begin{align}
    G&=K\,, &&  K=\begin{pmatrix} 0 & 1 \\ 1 & 0\end{pmatrix}\,,
\end{align}
which leads to 
\begin{align}
    S&= \langle e|R(\omega)\rangle+\langle \omega |R(e)\rangle\,.
\end{align}
For a single massless field in (A)dS$_3$ the action has exactly the same form, 
but the curvatures have a $\lambda^2$-correction. Now one can perform a linear 
change of variables and get
\begin{align}\label{eq:masslessgammaNdiag}
    \gamma_N&=\begin{pmatrix} 1 & 0 \\ 0 & -1\end{pmatrix}\,,
\end{align}
which leads to two decoupled actions, i.e., the matrix $K$ becomes 
numerically equal to $\gamma_N\,$.
Once the actions decouple, the relative coefficient can be made arbitrary, 
see e.g. \cite{Witten:1988hc,Blagojevic:2003vn, Peleteiro:2020ubv}.

\subsection{Strange higher spin systems}
\label{subsec:Strange}
We are now equipped with all the necessary machinery to generalize the examples of 
section \ref{sec:topological}. First, let us consider the topological system of 
coupled spin-two and -three fields in Minkowski space of section 
\ref{sec:topological}.  Indeed, from the gauge transformations 
\eqref{coupledgaugetrans} we see that the first derivative of one field enters the 
gauge transformations of the other. As done explicitly in section 
\ref{sec:firstorderreformulation} in the frame-like formulation, 
such terms can be obtained in two steps: 
1) one fixes the local Lorentz symmetry \eqref{gaugetransmassless} with parameter 
$\eta^{\ga(2s-2)}$ and the Lorentz gauge parameter $\eta^{\ga(2s-2)}$ gets expressed 
as the first derivative of the Fronsdal parameter $\xi^{\ga(2s-2)}$; 
2) the Lorentz gauge parameter $\eta^{\ga(2s-2)}$ enters the transformations of the 
vielbein of the other field in the system and the other way around. 

In spinor notation\footnote{The dictionary between vector and spinor notation 
is given in Appendix \ref{app:dico}.}, the gauge transformations of section 
\ref{sec:firstorderreformulation} are the form
\besubeqs
\begin{align}
    \delta e^{\ga\ga}&= d\eta^{\ga\ga}+h\fud{\ga}{\gb} \chi^{\ga\gb} +h_{\gb\gb} \rho^{\ga\ga\gb\gb}\,,\\
    \delta \omega^{\ga\ga}&= d\chi^{\ga\ga}\,,\\
    \delta e^{\ga\ga\ga\ga}&= d\xi^{\ga\ga\ga\ga}+h\fud{\ga}{\gb} \rho^{\ga\ga\ga\gb} +h^{\ga\ga} \chi^{\ga\ga}\,,\\
    \delta \omega^{\ga\ga\ga\ga}&= d\rho^{\ga\ga\ga\ga}\,.
\end{align}
\esubeqs
More generally, for any two neighbouring spins the following system 
is consistent
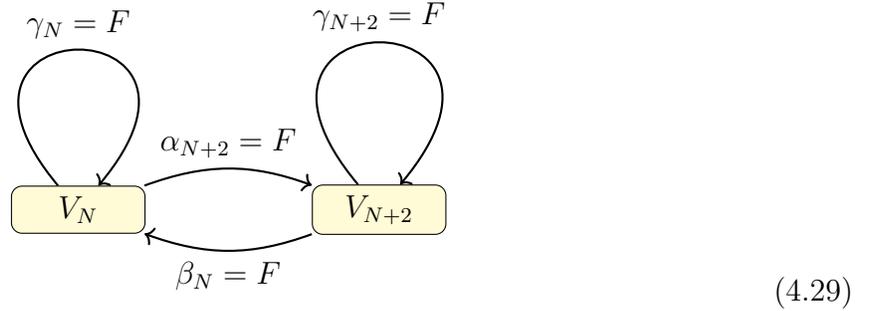
\begin{align}\label{twonodes}
\begin{tikzpicture}
    \filldraw[black] (0,0)  node[rectangle, rounded corners,draw,fill=yellow!20, text width=1.5cm, align=center] (B) {$V_N$};
    \filldraw[black] (4,0)  node[rectangle, rounded corners,draw,fill=yellow!20, text width=1.5cm, align=center] (C) {$V_{N+2}$};
    \draw[-latex,thick,->] (B) to[out=20,in=160] node[anchor=south] {$\alpha_{N+2}=F$} (C);
    \draw[-latex,thick,->] (C) to[out=-160,in=-20] node[anchor=north] {$\beta_N=F$} (B);
    \draw[-latex,thick,->] (B)  edge   [out=130,in=50,looseness=15]   node[anchor=south] {$\gamma_N=F$} (B);
    \draw[-latex,thick,->] (C)  edge   [out=130,in=50,looseness=15]   node[anchor=south] {$\gamma_{N+2}=F$} (C);
\end{tikzpicture}
\end{align}
In the case of section \ref{sec:topological} we need fields with $N=2,4$, but for any $N$ a representation of the quiver can be chosen to be
\begin{align}
    \gamma_N&=\gamma_{N+2}=\beta_N=F\,, &\alpha_{N+2}&= qF\,, && F=\begin{pmatrix} 0 & 1 \\ 0 & 0\end{pmatrix} .
\end{align}
These matrices are spin-independent, which is a particular solution. With the help 
of $GL(2)\times GL(2)$ transformations we can reach $\gamma_N=\gamma_{N+2}=\beta_N=F$, 
but $\alpha_{N+2}= q F$, where $q$ is a genuine parameter of this representation of 
the Poincar\'e algebra. Alternatively, we can choose 
$\gamma_N=\gamma_{N+2}=\alpha_{N+2}=F$ and $\beta_N= q F$.  Therefore, we have found a 
family of finite-dimensional representations of $iso(2,1)$ that depend on one free 
parameter.

Upon gauge fixing and going to the metric-like formalism one finds schematically
\begin{align}
    \delta \phi_s &= \pl \xi_{s-1} +\eta \epsilon \pl \xi_{s-2}\,,& 
    \delta \phi_{s-1}&= \pl\xi_{s-2} +\epsilon \pl \xi_{s-1}\,.
\end{align}
The field content, both frame-like and metric-like, matches the one required to 
describe a depth-$2$ partially-massless field. Indeed, there is a smooth deformation 
to (A)dS$_3$ with
\begin{align}&
\left(
\begin{array}{cc}
 0 & 1 \\
 \frac{(N+4)^2 \lambda ^2}{(N+2)^2} & 0 \\
\end{array}
\right)\,, && \left(
\begin{array}{cc}
 0 & 1 \\
 \frac{N^2\lambda ^2}{(N+2)^2} & 0 \\
\end{array}
\right)\,, && \left(
\begin{array}{cc}
 0 & 1 \\
 \frac{N(N+4) \lambda ^2}{2N^2} & 0 \\
\end{array}
\right)\,,&& \left(
\begin{array}{cc}
 0 & -\frac{4}{N(N+4)} \\
 -\frac{4\lambda ^2}{(N+2)^2} & 0 \\
\end{array}
\right)\,,
\end{align}
for the same matrices $\gamma_N\,,\gamma_{N+2}\,,\beta_N\,,\alpha_{N+2}$. 
Up to a simple linear $GL(2)\times GL(2)$ transformation of the fields the system is 
equivalent to the canonical form of the partially-massless system \eqref{canonicalPM}. 
There is no free parameter, of course. As it was mentioned around \eqref{diagcase}, 
the kinetic matrix can be chosen to be $N$-independent. In a different form the action 
can be found in \cite{Buchbinder:2012bz}.

Coming back to the case of $N=2,4$ of section \ref{sec:topological}, 
the existence of the various different systems found 
in (A)dS$_3$ can be understood from the following 
representation theory argument. 
Recall that $(j_1,j_2)$ denotes an irreducible representation of $so(2,2)$ or 
$sl(2,\mathbb{C})\cong so(1,3)$ of dimension $(2j_1+1)(2j_2+1)$ and $(j)$ denotes the 
dimension-$(2j+1)$ irreducible representation of the 
Lorentz subalgebra $sl_L(2,\mathbb{R})$.  
There are several ways to produce a one-form gauge field valued 
in a given Lorentz spin-$(j)$ representation 
within an (A)dS$_3$ one-form field $\omega^{\alpha(N),\dot{\alpha}(M)}\,$
transforming in the representation $(N/2,M/2)$ of the (A)dS$_3$ isometry algebra. 
Sticking to the case studied in section \ref{sec:topological}, 
there are many solutions since from the $sl_L(2,\mathbb{R})$ 
representations $2\times(1)$ and $2\times(2)$ 
corresponding to the generators $(P_a,J_a,P_{aa},J_{aa})$, 
one can think of various possible combinations among the set 
$\{ (1,0), (0,1), (2,0), (0,2), (3/2,1/2), (1/2,3/2)\}$ 
of (A)dS$_3$ representations.
Although the first four representations in this set are simple in their 
decomposition with respect to the Lorentz subalgebra, the latter two branch 
into the direct sum $(2)\oplus (1)\,$. 
The usage of the representations $(1,0)$ and $(2,0)$  
(or their conjugates) to represent two one-form gauge fields 
means that the corresponding spin-$2$ and spin-$3$ fields do not 
talk to each other, since they correspond to the two generators 
$(J_a,J_{aa})$ of the same $SL(3,\mathbb{R})$ group factor. 
As we already commented below Eq. \eqref{eq:abcdsolutionnomix}, 
in the free theory the 
spin-2 and spin-3 sectors do not mix in this situation.
One can also have systems without any straightforward metric-like 
interpretation, e.g. $(1,0)\oplus (2,0)\oplus (3/2,1/2)$, 
or $2\times (3/2,1/2)$.  
For example, the system \eqref{eq:soluA1} with $\eta = +1$ corresponds 
to $(3/2,1/2)\oplus (3/2,1/2)\,$.
The partially-massless systems in $(A)dS_3$ correspond to 
$(3/2,1/2)\oplus (1/2,3/2)$; the two decoupled systems to 
$(1,0)\oplus (2,0)$ and $(1,0)\oplus(0,2)$; the four $AdS_3$-models 
with one pair of fields decoupled correspond to 
$(3/2,1/2)\oplus X$, where $X$ is any 
combination of $(1,0)$ or $(0,1)$ with $(2,0)$ or $(0,2)$.

\paragraph{Possible interactions.}\label{interactions} Since any interacting
topological theory has to be of Chern-Simons form \cite{Grigoriev:2020lzu}, 
in order to find interactions we have to identify the fields associated  
with the generators of some Lie algebra that has a non-degenerate invariant 
bilinear form. 
Let us go back to the simplest case with spin-two and -three fields. 
After we moved to (A)dS$_3$ we are looking for an algebra that has two 
generators $T^{\ga(4)}$ and $T^{\ga(2)}$ 
(when decomposed into a sum of Lorentz modules) plus, possibly, additional 
generators associated to other fields that might be required to obtain a non-Abelian 
algebra. 
The two generators $T^{\ga(4)}$ and $T^{\ga(2)}$ can be understood as coming from a 
single partially-massless generator $T^{\ga(3),\gad(1)}$ and its conjugate 
$T^{\ga(1),\gad(3)}$ -- if we do not consider models where one pair of fields forms 
$(1,0)\oplus(2,0)$ and can be unified by one $sl(3,\mathbb{R})$.\footnote{We use here 
and below the notation $T^{\alpha(2j_1),\gad(2j_2)}$ to simply denote the 
corresponding representation of $sl(2,\mathbb{R})\oplus sl(2,\mathbb{R})$.}

One way to get a simple finite dimensional higher spin algebra\footnote{By a higher 
spin algebra we simply mean a Lie algebra that contains a given space-time symmetry 
algebra as a subalgebra and decomposes into irreducible modules that are larger than 
the adjoint ones, where the latter are associated with some higher spin fields. } is 
to take an irreducible module $V$ of the space-time symmetry algebra and consider 
$gl(V)=u(1)\oplus sl(V)$. 
One can also apply this construction to a module 
$V$ that is a direct sum of irreducible modules. 
In order to get the required spectrum 
from a higher spin algebra of $gl(V)$-type we can take 
$V= T^{\ga(2)} \oplus T^{\ga,\gad}$ of the (anti)-de Sitter algebra, i.e.
we cannot consider just a single irreducible representation of $sl(2,\mathbb{R})$. 
The full spectrum is then given by the tensor product 
$V\otimes V^*$ and reads   
$2\times T^{\alpha(2),\dot{\alpha}(2)} \oplus T^{\alpha(3),\dot{\alpha}}\oplus 
T^{\alpha,\dot{\alpha}(3)}\oplus 2\times T^{\alpha,\dot{\alpha}}\oplus 
T^{\alpha(2)}\oplus T^{\dot{\alpha}(2)}\oplus T$.
This seems to be the most minimal extension that admits interactions 
and contains the spin-two subsector.  
It is clear that there is no algebra that contains 
only the generators $T^{\ga(3),\gad(1)}\oplus T^{\ga(1),\gad(3)}$, 
which a posteriori explains the no-go result of 
section \ref{sec:topological}. Indeed, there is no $V$ such that 
$V\otimes V^*$ gives just $(3/2,1/2)\oplus(1/2,3/2)\oplus(0,0)$, 
which can be seen by enumerating a handful of low-spin representations $V$. 
We postpone to a future work the analysis of the model based 
on these fields in AdS$_3$ and their flat limit.

\paragraph{Simple generalization.} The system above has an obvious generalization 
to a topological system that covers a range of spins. We can extend the system by 
duplicating the nodes and defining $\alpha\sim \beta\sim\gamma \sim F$. It will always 
be consistent since $FF=0$. The (first few levels of the) gauge transformations look 
schematically as
\besubeqs
\begin{align}
    \delta \phi_s &= \pl \xi_{s-1} +\eta \epsilon \pl \xi_{s-2}\,,\\
    \delta \phi_{s-1}&= \pl\xi_{s-2} +\epsilon \pl \xi_{s-1}+ \eta \epsilon \pl 
    \xi_{s-3}\,,\\
    \delta \phi_{s-2}&= \pl \xi_{s-3}+\epsilon \pl \xi_{s-2}+\eta \epsilon \pl 
    \xi_{s-4}\,,
\end{align}
\esubeqs
and can extend down to any spin $s\geq 1$. This system has more free parameters: 
there is one parameter per each $\alpha$ (or $\beta$), which gives other examples 
of finite-dimensional representations of $iso(2,1)$.  

An (A)dS$_3$-deformation of such a system is a partially-massless field that 
originates from $T^{\ga(2s-2-k),\gad(k)}$ and $T^{\ga(k),\gad(2s-2-k)}$. After 
decomposing with respect to the diagonal Lorentz algebra, its top spin component is 
$T^{\ga(2s-2)}$ and the lowest one is $T^{\ga(2s-2-2k)}$. For the same reason as 
before, this system does not admit interactions unless we extend it with more fields, 
assuming the deformation of interactions has to be smooth in the cosmological constant.

\paragraph{Even stranger systems.} Another interesting example is a system 
that contains fields of spins $2$, $2$, $3$, $4$ or, more generally, 
$s$, $s$, $s+1$, $s+2$. It corresponds to the following quiver
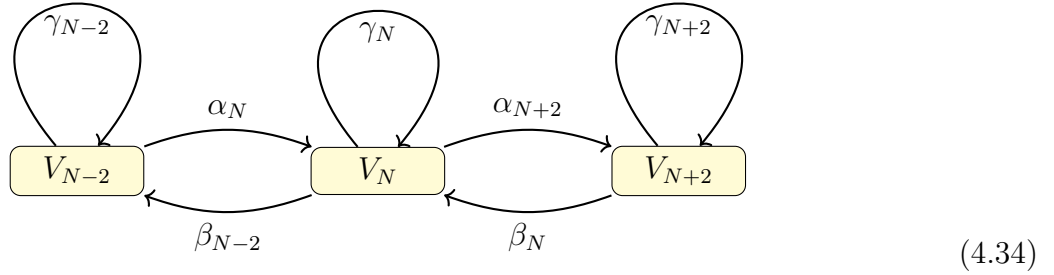
\begin{align}
\begin{tikzpicture}
    \filldraw[black] (-4,0)  node[rectangle, rounded corners,draw,fill=yellow!20, text width=1.5cm, align=center] (A) {$V_{N-2}$};
    \filldraw[black] (0,0)  node[rectangle, rounded corners,draw,fill=yellow!20, text width=1.5cm, align=center] (B) {$V_N$};
    \filldraw[black] (4,0)  node[rectangle, rounded corners,draw,fill=yellow!20, text width=1.5cm, align=center] (C) {$V_{N+2}$};
    \draw[-latex,thick,->] (B) to[out=-160,in=-20] node[anchor=north] {$\beta_{N-2}$} (A);
    \draw[-latex,thick,->] (A) to[out=20,in=160] node[anchor=south] {$\alpha_N$} (B);
    \draw[-latex,thick,->] (B) to[out=20,in=160] node[anchor=south] {$\alpha_{N+2}$} (C);
    \draw[-latex,thick,->] (C) to[out=-160,in=-20] node[anchor=north] {$\beta_N$} (B);
    \draw[-latex,thick,->] (B)  edge   [out=130,in=50,looseness=15]   node[anchor=north] {$\gamma_N$} (B);
    \draw[-latex,thick,->] (A)  edge   [out=130,in=50,looseness=15]   node[anchor=north] {$\gamma_{N-2}$} (A);
    \draw[-latex,thick,->] (C)  edge   [out=130,in=50,looseness=15]   node[anchor=north] {$\gamma_{N+2}$} (C);
\end{tikzpicture}
\end{align}
with a representation given by
\begin{align}
    \alpha_{N+2}=\beta_N=\gamma_{N+2}=\gamma_N=F\,,\\
    \gamma_{N-2}=\begin{pmatrix} F & 0 \\ 0 & F\end{pmatrix}\,, \qquad \alpha_N=\begin{pmatrix} F & F \end{pmatrix}\,, \qquad \beta_{N-2}=\begin{pmatrix} F \\ F\end{pmatrix} \,.
\end{align}
Here-above, we provide an obvious generalization to any spin. 
The action can be written with
\begin{align}
    G_{N-2}&=\begin{pmatrix} K & 0 \\ 0 & K\end{pmatrix}\,, && -G_N=G_{N+2}=K\,.
\end{align}
The AdS deformation leads to $T^{\ga(4),\gad(2)}$ and $T^{\ga(2)}$, i.e. it is a 
partially-massless field and a massless one that are decoupled from each other. 

It is easy to generalize this example to more complicated systems. We can begin with 
any number $k$ of $V_{N-2i}$, $i=0,...,k$ that are even dimensional. $\gamma_{N-2i}$ 
can be block diagonal made of $F$, and $\alpha$, $\beta$ can mix them. Nilpotency of 
$F$ ensures that the system is consistent. Some free parameters can be introduced in 
the same way as before.

\section{Conclusions}

In this paper, we studied some higher-spin systems in three-dimensional 
Minkowski space originally found in \cite{Boulanger:2020yib}, following the 
higher off-shell dualisation procedure proposed originally 
in \cite{Boulanger:2012df}.
We found several generalisations of these systems in flat space. 
Firstly, we found that these strange higher-spin actions in flat space 
admit a one-parameter extension. 
Secondly, we found that for some discrete values of the parameter in 
the action, these system could be deformed to the (A)dS$_3$ background. 
Thirdly, we found various generalisations of these models to larger spectra 
of fields having spin even higher than three, both in flat and (A)dS$_3$.

At the free level, an interesting mathematical problem we have encountered is the 
classification of finite-dimensional representations of Poincar\'e algebra, 
since each of such representations defines one of our free topological systems.
To the best of our knowledge this problem remains unsolved. 

Concerning possible interactions at the action level, 
once the flat space system is assumed to have a smooth deformation to (A)dS$_3$ 
the powerful theorems on the representation theory of (semi)-simple Lie algebras 
are at our disposal. 
Free topological systems in (A)dS$_3$ are simpler to study as compared 
to topological systems in flat space, since we know all the finite-dimensional 
representations of the (A)dS$_3$ isometry algebras.
As we discuss in section \ref{subsec:Strange}, 
all metric-like topological (A)dS$_3$ systems contain 
(partially)-massless fields and nothing else.
Then, as far as interactions are concerned, the (A)dS$_3$ background
also makes the search for interactions simpler, since we know which 
spectrum of fields to introduce in order to have a $gl(V)$ associative 
matrix algebra, out of which a Lie algebra is obtained by taking the 
commutator, the trace operation being the trace of matrices in $End(V)$.
It remains to be seen if there are genuine 
interacting topological theories in flat space, i.e. those that do not admit 
any deformations to (A)dS$_3$.

Since we found some (A)dS$_3$ models that are not analytical in the cosmological 
constant, it is not yet clear whether there 
could be some non-Abelian theory in (A)dS$_3$ that would be non-analytical 
in the cosmological constant, hence admitting no flat limit. 
We leave this for future investigations. 
The simplest spin-2/spin-3 models \eqref{eq:simplesol23} -- \eqref{G0H0} 
that we found in (A)dS$_3$ 
admit a smooth flat limit and therefore cannot allow for 
a non-Abelian deformation that would be analytical in the cosmological constant.

As another possible outlook, it would be interesting to look at the asymptotic 
symmetries of the new topological higher-spin systems we found. 
In AdS such analysis should fit within the extension of the  asymptotics 
of massless higher-spin fields \cite{Henneaux:2010xg, Campoleoni:2010zq, Campoleoni:2011hg}  modulo possible generalisations of the ``standard'' 
boundary conditions along the lines, e.g., of \cite{Grumiller:2016pqb}. 
On the other hand, in flat space the variety of inequivalent bulk symplectic 
structure that we identified should lead to a rich landscape of higher-spin 
asymptotic symmetries beyond those discussed, e.g., in \cite{Afshar:2013vka, Gonzalez:2013oaa, Campoleoni:2016vsh, Ammon:2017vwt} and references therein. 

\subsection*{Acknowledgements}

We are grateful to Antoine Bourget for useful discussions 
on quiver representations. We would also like to thank 
Gaston Giribet, Stam Nicolis and Massimo Porrati for discussions.

AC and ES are research associates of the Fund for Scientific Research – FNRS, Belgium. 
This work was partially supported by the FNRS through the grants No. FC.36447, 
No. F.4503.20, and No. T.0022.19. 
The work of ES was supported by the European Research Council (ERC) under the 
European Union’s Horizon 2020 research and innovation programme (grant agreement 
No 101002551). The work of VL was funded by the European Union’s Horizon 2020 
research and innovation programme under the Marie Skłodowska-Curie grant 
agreement No 101034383.

\appendix

\section{Dictionary between spinor and vector notation}
\label{app:dico}

We introduce a basis of three real, symmetric, $2\times 2$ matrices 
$\tau^a = (\tau^a_{\alpha\beta})=(\tau^a_{\beta\alpha})$ and raise (lower) 
the indices according to 
$q^\alpha = \epsilon^{\alpha\beta}\,q_{\beta} 
=\epsilon^{\alpha\beta}\,q^{\gamma}\epsilon_{\gamma\beta}\,$.
The three $\tau^a$ matrices obey the orthogonality and completeness relations
\begin{equation}
    \tau^a{}_{\alpha\beta}\,\tau^b{}^{\alpha\beta} = -2\,\eta^{ab}\,,
    \quad 
    \tau^a{}_{\alpha\beta}\,\tau_a{}^{\gamma\delta} = 
    -2\,\delta_{(\alpha}^\gamma\delta_{\beta)}^\delta\,,
\end{equation}
where we use the mostly plus convention $(\eta_{ab})=\;$diag$(-1,+1,+1)\,$
together with 
\begin{equation}
    \tau^a_{\alpha\beta}\,\tau^b{}^{\beta\gamma} 
    = -\eta^{ab}\,\delta^\gamma_\alpha + \epsilon^{abc}\,\tau_{c\,\alpha}{}^\gamma\;.
\end{equation}
In particular, one has
\begin{equation}
    \epsilon^{abc} = -\tfrac{1}{2}\,{\rm Tr}(\tau^a\tau^b\tau^c)\;,
    \qquad {\rm where}\qquad \epsilon^{012}=1\;. 
\end{equation}
The dictionary between vector and spinor notation is
\begin{equation}
    V^a = \tau^a_{\alpha\beta}\,V^{\alpha\beta}\qquad\Leftrightarrow\qquad 
    V^{\alpha\beta} = -\tfrac{1}{2}\,\tau_a^{\alpha\beta}\,V^a\;.
\end{equation}
Therefore, associated with the transformation laws in vector notation
\begin{align}
    \delta e^a &= {\rm d} \xi^a + \alpha\,\epsilon^{abc}\,h_b\,\Lambda_c
    +\beta\,h_b\,\alpha^{ab}\;,
    \\
    \delta e^{aa} &= {\rm d} \xi^{aa} + \gamma\,h_b\,\epsilon^{abc}\,\alpha_c{}^a
    +\sigma\,(h^a\Lambda^a - \tfrac{1}{3}\,\eta^{aa}\,h_b\,\Lambda^b)\;,
\end{align}
one has, respectively, the following transformations
\begin{align}
    \delta e^{\alpha\alpha} &= {\rm d} \xi^{\alpha\alpha} + 
    2\,\alpha\,h^{\beta\alpha}\,\Lambda^{\alpha}{}_{\beta} 
    - 2\,\beta\,h_{\beta\beta}\,\alpha^{\alpha\alpha\beta\beta}\;,
    \\
    \delta e^{\alpha(4)} &= {\rm d} \xi^{\alpha(4)} + 
    2\,\gamma\,h^{\beta\alpha}\,\alpha_{\beta}{}^{\alpha(3)} 
    + \sigma\,h^{\alpha\alpha}\,\Lambda^{\alpha\alpha}\;.
\end{align}

From the definition of the operator $Q\,$ in \eqref{eq:Q}, we have
\begin{align}
    \delta e^{\alpha\alpha} &= \nabla \xi^{\alpha\alpha} + 
    \beta(2)_{12}\, h_{\beta\beta}\rho^{\alpha\alpha\beta\beta}
    + 2\, \gamma(2)_{12} h_{\beta}{}^\alpha \chi^{\alpha\beta}
    \\
    \delta \omega^{\alpha\alpha} &= \nabla \chi^{\alpha\alpha} + 
    \beta(2)_{21}\, h_{\beta\beta}\xi^{\alpha\alpha\beta\beta}
    + 2\, \gamma(2)_{21} h_{\beta}{}^\alpha \xi^{\alpha\beta}
    \\
    \delta e^{\alpha(4)} &= \nabla \xi^{\alpha(4)} + 
    12\,\alpha(4)_{12}\, h^{\alpha\alpha}\chi^{\alpha\alpha}
    + 4 \,\gamma(4)_{12} h_{\beta}{}^\alpha \rho^{\alpha(3)\beta}
    \\
    \delta \omega^{\alpha(4)} &= \nabla \rho^{\alpha(4)} + 
    12\,\alpha(4)_{21}\, h^{\alpha\alpha}\xi^{\alpha\alpha}
    + 4 \,\gamma(4)_{21} h_{\beta}{}^\alpha \xi^{\alpha(3)\beta}\;
\end{align}
for some $2\times 2$ matrices $\beta(2)\,$, $\gamma(2)\,$, $\gamma(4)\,$ 
and $\alpha(4)\,$. 

On the other hand, doing the translation between vector and spinor notation, 
we find that, associated with the transformation laws in spinor notation
\begin{align}
    \delta e^{\alpha\alpha} &= {\rm d} \xi^{\alpha\alpha} + 
    2\,\alpha\,h^{\beta\alpha}\,\Lambda^{\alpha}{}_{\beta} 
    - 2\,\beta\,h_{\beta\beta}\,\alpha^{\alpha\alpha\beta\beta}\;,
    \\
    \delta e^{\alpha(4)} &= {\rm d} \xi^{\alpha(4)} + 
    2\,\gamma\,h^{\beta\alpha}\,\alpha_{\beta}{}^{\alpha(3)} 
    + \sigma\,h^{\alpha\alpha}\,\Lambda^{\alpha\alpha}\;,
\end{align}
there corresponds the following transformations in the vector notation:
\begin{align}
    \delta e^a &= {\rm d} \xi^a + \alpha\,\epsilon^{abc}\,h_b\,\Lambda_c
    +\beta\,h_b\,\alpha^{ab}\;,
    \\
    \delta e^{aa} &= {\rm d} \xi^{aa} + \gamma\,h_b\,\epsilon^{abc}\,\alpha_c{}^a
    +\sigma\,(h^a\Lambda^a - \tfrac{1}{3}\,\eta^{aa}\,h_b\,\Lambda^b)\;.
\end{align}
From the above dictionary and \eqref{eq:gauge23matrix},  
we find the following identification of $2\times 2$ matrices:
\begin{equation}
    \gamma(2)=-A\, , \quad 2\gamma(4)=-D\, , \quad 12\alpha(4) = C\, , 
    \quad \beta(2) = - 2 B\, .
\end{equation}
Then the equations \eqref{eq:matrixeqs} and \eqref{integrability} 
are in perfect agreement.
With the further identification
\begin{equation}
    G = - 2 G(2)\; , \quad H =  \tfrac{1}{12} G(4)\; ,
\end{equation}
equations \eqref{eq:actionmetric23} and \eqref{eq:actionmetricgeneral} agree as well.

\section{Some definitions about quivers}
\label{app:quivers}
We recall verbatim from \cite{kirillov2016quiver} some definitions 
and results about quivers and their representations 
that we refer to in the main body of the paper:
\begin{itemize}
    \item A quiver $\vec{Q}$ is a directed graph; 
formally it can be described by a set of vertices $I\,$, 
a set of edges $\Omega\,$, and two maps $s,t:\Omega\rightarrow I$
which assign to every edge its source and target, respectively.
One can also think of a quiver $\vec{Q}$ as a graph $Q$ along 
with an orientation, i.e., choosing for each edge of $Q\,$, which 
of the two endpoints is the source and which is the target. 
It is assumed that the set of edges and vertices are finite and 
that $\vec{Q}$ is connected. 
A representation of a quiver $\vec{Q}$ is the following collection 
of data:
\begin{itemize}
    \item For every vertex $i\in I\,$,  vector space $V_i$ over the 
    field $\mathbb{K}\,$;
    \item For every edge $h\in \Omega\,$, $h: i\rightarrow j\,$, 
    a linear operator $x_h: V_i\rightarrow V_j\,$ .
\end{itemize}
For the quivers considered in this paper, the operators $x_h$ also have 
to satisfy certain quadratic relations such as \eqref{eq:matrixeqs} or 
\eqref{integrability}.
A simple example of a quiver is given by the Jordan quiver: 
\begin{align}
\vec{Q}_J = \xymatrix{
 \bullet \ar@(ur,dr)[]}
 \end{align}
and a representation of this quiver is the pair $(V,x)$ where 
$V$ is a vector space over the field $\mathbb{K}$ and $x:V\rightarrow V$
is a linear map. Classifying the representations of $\vec{Q}_J$
is equivalent to classifying matrices up to similarity.

\item A quiver $\vec{Q}$ is called \emph{Dynkin} iif the underlying graph $Q$
is one of the following graphs:
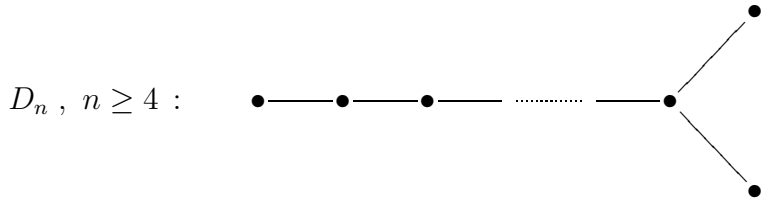
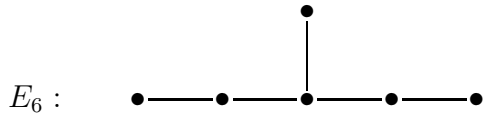
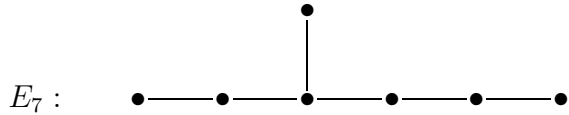
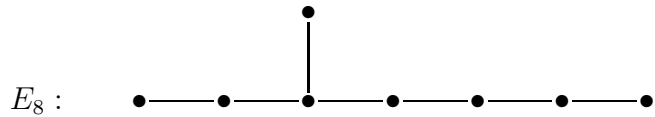
\begin{align}\nonumber
\xymatrix@1{A_n\;,~n\geq 1\,: & 
 \bullet\ar@{-}[r]& \bullet\ar@{-}[r]& \bullet\ar@{-}[r] 
 & ~\ar@{..}[r] & ~\ar@{-}[r]& \bullet\ar@{-}[r]& \bullet & & }
 \nonumber\\
\xymatrix@1{  &  &  &  &  &  &  & \bullet & & 
\nonumber\\
D_n\;,~n\geq 4\,: & 
 \bullet\ar@{-}[r]& \bullet\ar@{-}[r]& \bullet\ar@{-}[r] & 
 ~\ar@{..}[r] & ~\ar@{-}[r]& \bullet\ar@{-}[ur]\ar@{-}[dr] & & & 
 \nonumber\\
 &  &  &  &  &  &  & \bullet & & }
 \nonumber\\
 \xymatrix@1{ & & & &  &  & \bullet\ar@{-}[d]  &  & & & & &
 \nonumber\\
 & & & E_6:&\bullet\ar@{-}[r]& \bullet\ar@{-}[r]& \bullet\ar@{-}[r]
 & \bullet\ar@{-}[r] & \bullet & & & &}
 \nonumber\\
 \xymatrix@1{ & & & & &  &  & \bullet\ar@{-}[d]  &  & & & & &
 \nonumber\\
 & & & & E_7:&  \bullet\ar@{-}[r]& \bullet\ar@{-}[r]& \bullet\ar@{-}[r]
 & \bullet\ar@{-}[r] & \bullet\ar@{-}[r] & \bullet & & &}
 \nonumber\\
 \xymatrix@1{ & & & & &  &  & \bullet\ar@{-}[d]  &  & & & & &
 \nonumber\\
 & & & & E_8: & \bullet\ar@{-}[r]& \bullet\ar@{-}[r]& \bullet\ar@{-}[r]
 & \bullet\ar@{-}[r] & \bullet\ar@{-}[r] & \bullet\ar@{-}[r] & \bullet & &}
 \nonumber
\end{align}
These are the Dynkin diagrams of simply-laced, finite-dimensional, 
simple Lie algebras over the complex numbers.

\item A quiver $\vec{Q}$ is called \emph{Euclidean} iif the underlying graph $Q$
is one of the following graphs:
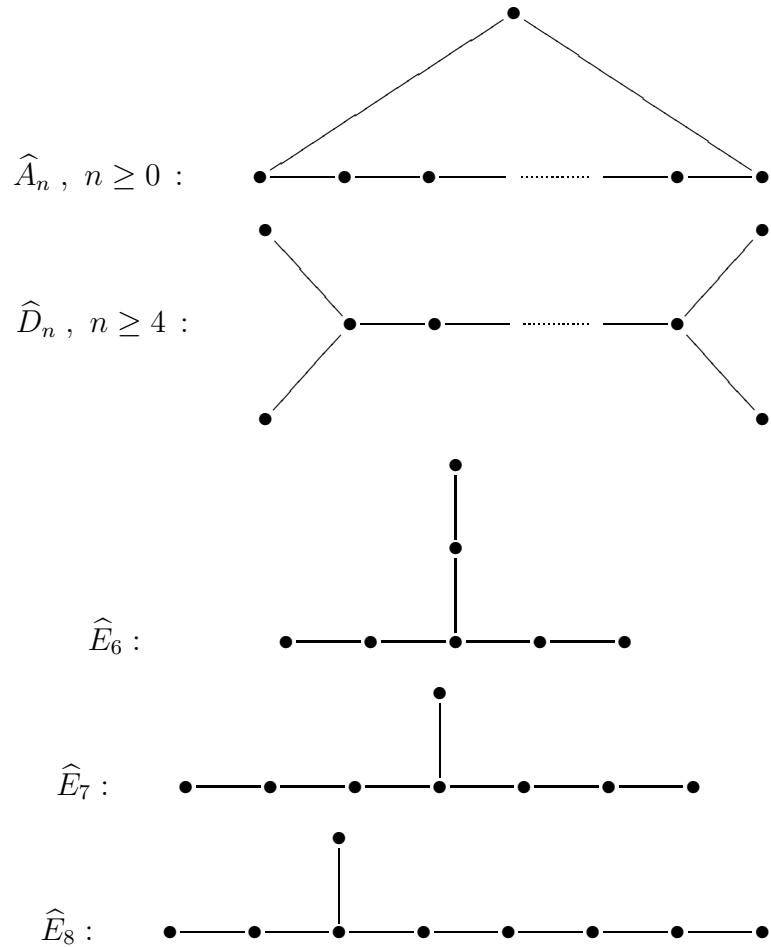
\begin{align}
\nonumber
\xymatrix@1{ &  & & & \bullet\ar@{-}[ddrrr] & & & & & &  
\\
& & & & & & & & & & \nonumber \\
\widehat{A}_n\;,~n\geq 0\,: & \bullet\ar@{-}[r]\ar@{-}[uurrr]
& \bullet\ar@{-}[r]& \bullet\ar@{-}[r] & ~\ar@{..}[r] 
& ~\ar@{-}[r]& \bullet\ar@{-}[r]& \bullet & & &}
 \nonumber \\
\xymatrix@1{  & \bullet\ar@{-}[dr] &  &  &  &  &  & \bullet & & &
\nonumber \\
\widehat{D}_n\;,~n\geq 4\,: & 
 & \bullet\ar@{-}[r]& \bullet\ar@{-}[r] & 
 ~\ar@{..}[r] & ~\ar@{-}[r]& \bullet\ar@{-}[ur]\ar@{-}[dr] & & & &
 \nonumber\\
 & \bullet\ar@{-}[ur]  &  &  &  &  &  & \bullet & & &}
 \nonumber \\
 \xymatrix@1{ & & & &  &  & \bullet\ar@{-}[d]  &  & & & & & &
 \nonumber \\
 & &  & &  &  & \bullet\ar@{-}[d]  &  & & & & & &
 \nonumber\\
 & & \widehat{E}_6: & &\bullet\ar@{-}[r]& \bullet\ar@{-}[r]& \bullet\ar@{-}[r]
 & \bullet\ar@{-}[r] & \bullet & & & & &}
 \nonumber\\
 \xymatrix@1{& & & & & &  &  & \bullet\ar@{-}[d]  &  & & & & & &
 \nonumber\\
 & & & & \widehat{E}_7:&  \bullet\ar@{-}[r]&\bullet\ar@{-}[r]
 & \bullet\ar@{-}[r]& \bullet\ar@{-}[r]
 & \bullet\ar@{-}[r] & \bullet\ar@{-}[r] & \bullet & & & &}
 \nonumber\\
 \xymatrix@1{ & & & & &  &  & \bullet\ar@{-}[d]  &  & & & & & & &
 \nonumber \\
 & & & & \widehat{E}_8: & \bullet\ar@{-}[r]& \bullet\ar@{-}[r]& \bullet\ar@{-}[r]
 & \bullet\ar@{-}[r] & \bullet\ar@{-}[r] & \bullet\ar@{-}[r] & \bullet
 \ar@{-}[r] & \bullet & & &}
\end{align}
These are the Dynkin diagrams of simply-laced affine Kac-Moody algebras.
\end{itemize}
The following results are quoted in \cite{kirillov2016quiver}, with 
references to the proofs given therein:

\begin{itemize}
    \item Quivers can be \emph{tame} or \emph{wild}; formal definitions of these concepts can be found in chapter 7 of \cite{kirillov2016quiver} and will not be reproced here. However, a simple characterization holds 
    \cite[theorem 7.47]{kirillov2016quiver}:
    
    \textbf{Theorem:} Let $\vec{Q}$ be a  connected quiver.
    \begin{enumerate}
        \item[(1)] If $\vec{Q}$ is Dynkin or Euclidean, then it is tame;
        \item[(2)] If $\vec{Q}$ is neither Dynkin nor Euclidean, then it is wild.
    \end{enumerate}

    Representations of tame quivers have been classified. On the other hand, the representations of wild quivers are not known in general. The quivers considered in this paper are wild.
\end{itemize}


\providecommand{\href}[2]{#2}\begingroup\raggedright\endgroup

\end{document}